\documentclass[onecolumn,11pt]{article}
\usepackage[top=1in, bottom=1in, left=1in, right=1in]{geometry}
\usepackage{amsmath,amsfonts,amscd,amssymb}
\usepackage{graphicx}
\usepackage{epstopdf}
\usepackage{overpic}
\usepackage{cancel}
\usepackage{rotating}
\usepackage{url}
\usepackage{caption}
\usepackage{color}
\usepackage{rotating}
\usepackage{multirow}
\usepackage{wrapfig}
\usepackage{sidecap}
\usepackage{mathtools}
\usepackage{bbm}
\usepackage{subeqnarray}
\usepackage{setspace}
\usepackage{palatino} 
\usepackage{physics}
\usepackage[bottom,flushmargin,hang,multiple]{footmisc}
\usepackage{lipsum}
\newcommand\blfootnote[1]{%
  \begingroup
  \renewcommand\thefootnote{}\footnote{#1}%
  \addtocounter{footnote}{-1}%
  \endgroup
}

\usepackage[numbers,sort&compress]{natbib}

\definecolor{offlinered}{RGB}{115, 0, 0}
\definecolor{onlineblue}{RGB}{4, 79, 149}

\newcommand{\R}{\rm I\!R}
\renewcommand{\Re}{\textrm{Re}}

\title{\LARGE{\vspace{-.575in}\textbf{
Learning dominant physical processes \\ with data-driven balance models}}
		\vspace{-.15in}}
	\author{Jared L. Callaham$^{1*}$, James V. Koch$^{2}$, Bingni W. Brunton$^3$,\\ J. Nathan Kutz$^4$, and Steven L. Brunton$^1$\\
	\footnotesize{$^1$ Department of Mechanical Engineering, University of Washington, Seattle, WA 98195, United States}\vspace{-.05in}\\
	\footnotesize{$^2$ Oden Institute for Computational \& Engineering Sciences, University of Texas, Austin, TX 78712}\vspace{-.05in}\\
	\footnotesize{$^3$ Department of Biology, University of Washington, Seattle, WA 98195, United States}\vspace{-.05in}\\ 
	\footnotesize{$^4$ Department of Applied Mathematics, University of Washington, Seattle, WA 98195, United States}\vspace{-.05in}\\
	\footnotesize{$^*$ Correspondence to: jc244@uw.edu}\vspace{-.175in}
}
\date{}
\begin{document}
\maketitle

\blfootnote{$^*$ Corresponding author (jc244@uw.edu)\\ \noindent \textbf{Python code:}  github.com/dynamicslab/dominant-balance}

\vspace{-.2in}
\begin{abstract}
Throughout the history of science, physics-based modeling has relied on judiciously approximating observed dynamics as a balance between a few dominant processes.  
However, this traditional approach is mathematically cumbersome and only applies in asymptotic regimes where there is a strict separation of scales in the physics.  
Here, we automate and generalize this approach to non-asymptotic regimes by introducing the idea of an equation space, in which different local balances appear as distinct subspace clusters.
Unsupervised learning can then automatically identify regions where groups of terms may be neglected. 
We show that our data-driven balance models successfully delineate dominant balance physics in a much richer class of systems.  
In particular, this approach uncovers key mechanistic models in turbulence,  combustion, nonlinear optics, geophysical fluids, and neuroscience.  \\
\vspace{-0.05in}


\end{abstract}

\section{Introduction}\label{sec:introduction}

It is well known across the engineering and physical sciences that persistent behaviors in complex systems are often determined by the balance of just a few dominant physical processes.
This heuristic, which we refer to as dominant balance, has played a pivotal role in our study of systems as diverse as turbulence~\cite{Holmes1996}, geophysical fluid dynamics~\cite{Gill1982book,Lighthill1966jfm}, and fiber optics~\cite{Blow1989}.
It is also thought to play a role in the emerging fields of pattern formation~\cite{Cross1993,Morris1993prl,Grzybowski2000nature}, wrinkling~\cite{Cerda2003prl}, droplet formation~\cite{Shi1994science}, and biofilm dynamics~\cite{Seminara2012pnas}.
These balance relations provide reduced-order mechanistic models to approximate the full complexity of the system with a tractable subset of the physics.  

The success of dominant balance models is particularly evident in the field of fluid mechanics.  
The Navier-Stokes equations describe behavior across a tremendous range of scales, from water droplets to supersonic aircraft and hurricanes.  
Thus, much of our progress has required simplifying the physics with nondimensional parameters that determine which terms are important for a specific problem.  
Perhaps the most well-known dimensionless quantity, the Reynolds number, embodies the balance between inertial and viscous forces in a fluid. 
Other nondimensional numbers capture the relative importance of inertial and Coriolis forces (Rossby number), inertia and buoyancy (Froude number), and thermal diffusion and convection (Rayleigh number), among dozens of other possible effects.
In many situations, the magnitude of these coefficients determine the important mechanisms at work in a flow; conversely, they determine which mechanisms may be safely neglected.  
In geophysical flows, balance arguments bypass the incredible complexity of the ocean and atmosphere to identify driving mechanisms such as geostrophy, the thermal wind, Ekman layers, and western boundary currents~\cite{Lighthill1966jfm,Gill1982book}.
Lighthill, one of the most influential fluid dynamicists of the 20th century, often relied on dominant balance arguments as physical motivation for his mathematical analyses~\cite{Lighthill1952prs,Lighthill1966jfm}.
Beyond fluid mechanics, asymptotic methods have been crucial in characterizing a diverse range of physical behavior.  

Advanced statistical tools now allow analysis of the increasing wealth of data from modern experimental and numerical methods, but to date there is no direct link between this data and the powerful insights of asymptotic scaling analysis.
This presents an exciting opportunity to leverage data-driven methods, which are driving changes in a wide range of fields, from control \cite{Pastoor2008jfm, Verma2018pnas} to turbulence modeling \cite{Duraisamy2018arfm}, forecasting \cite{Lguensat2017}, and extreme event prediction \cite{Wan2018plos}.
Although some studies have addressed the dominant balance problem by using expert knowledge to design application-specific clustering algorithms~\cite{Portwood2016jfm,Lee2018} or a post hoc interpretation of unsupervised clustering in terms of dominant balance~\cite{Sonnewald2019ess}, to our knowledge the general challenge of identifying local dominant balance regimes directly from data remains open.

\begin{SCfigure*}
	\begin{overpic}[width=0.6\linewidth]{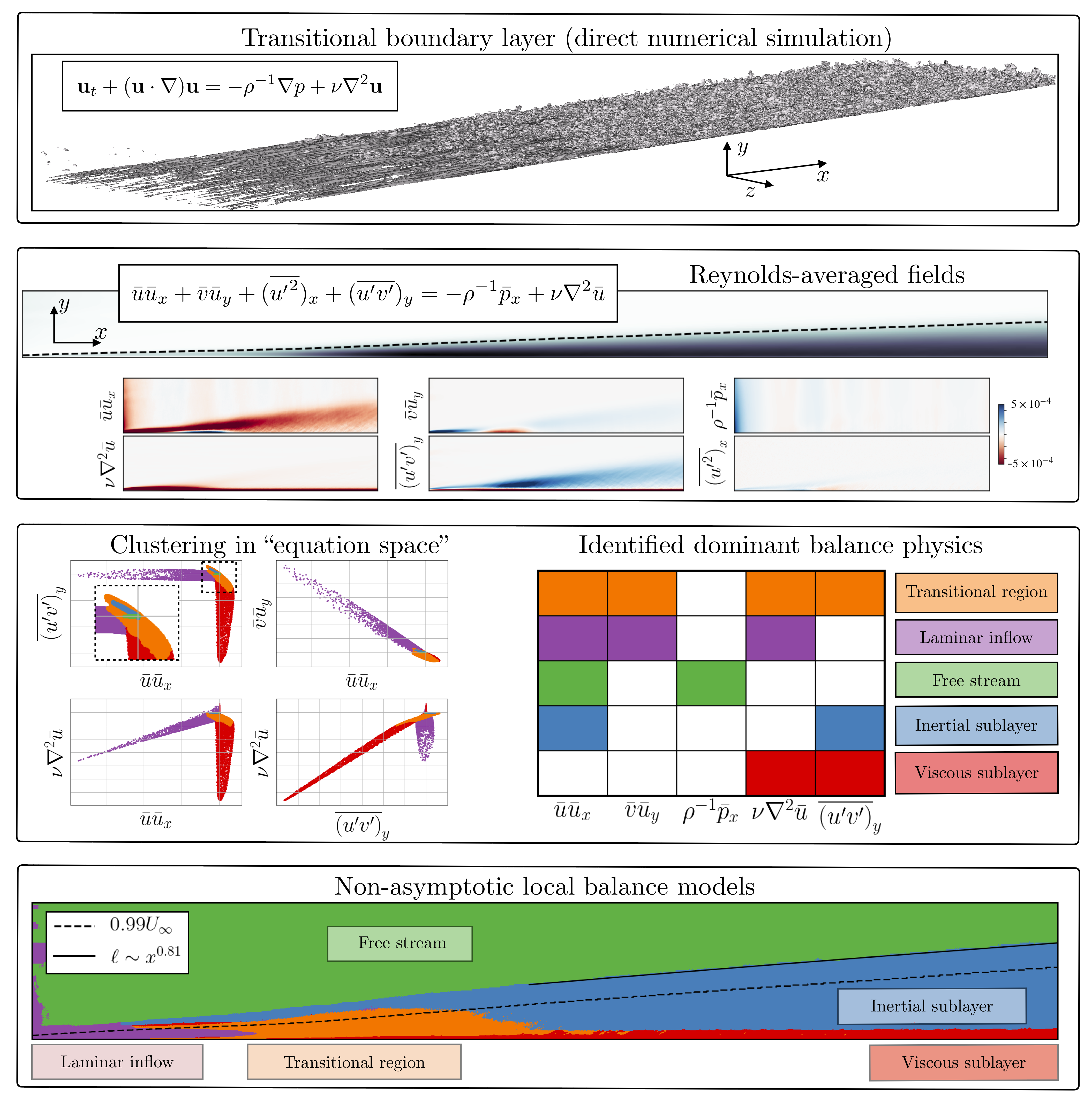}
		\footnotesize{
			\put(1.8, 96.5){\textbf{a}}
			\put(1.8, 75){\textbf{b}}
			\put(1.8, 50){\textbf{c}}
			\put(1.8, 18.8){\textbf{d}}
		}
	\end{overpic}
	\vspace{-.15in}
	\caption{ \small{
			\textbf{Schematic of the dominant balance identification procedure applied to a turbulent boundary layer.}
			High-resolution direct numerical simulation results (\textbf{a}, visualized with a turbulent kinetic energy isosurface) are averaged to compute the terms in the Reynolds-averaged Navier-Stokes equations (\textbf{b}).
			The equation space representation of the field enables clustering and sparse approximation methods to extract the distinct geometrical structures in the six-dimensional space corresponding to dominant balance physics (\textbf{c}). 
			Finally, the entire domain can be segmented according to these interpretable balance models, identifying distinct physical regimes (\textbf{d}).
			A curve fit to the wall-normal extent of the post-transition region of the identified inertial sublayer shows an approximate scaling of $ \ell \sim x^{0.81} $, consistent with the theoretical prediction of $ x^{4/5} $ from boundary layer theory.
			The 99~\% free-stream velocity ($U_\infty$) contour is also shown for reference.}
	}
	\label{fig:boundary-layer}
	\vspace{0.3in}
\end{SCfigure*}
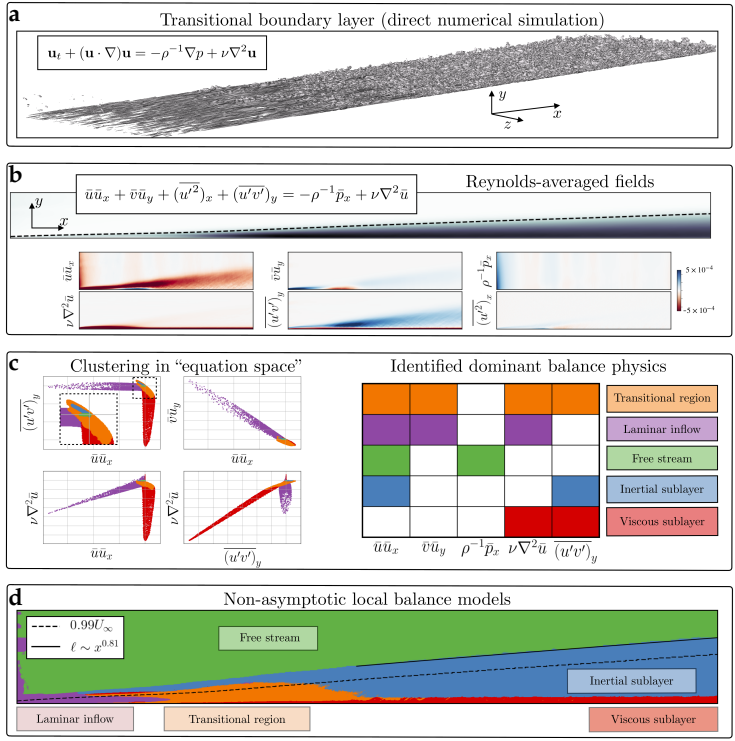

In this work, we develop a generalized data-driven method to identify dominant balance regimes in complex physical systems.
Figure~\ref{fig:boundary-layer} demonstrates the method applied to fluid flow over a flat plate in transition to turbulence.
We introduce a geometric perspective on dominant balance in which standard machine learning tools can automatically identify dominant physical processes.
The geometric approach naturally links the analysis to the underlying equation so that the entire procedure can be easily interpreted and visualized.
This data-driven method is designed to be applied in tandem with, rather than supplant, classical asymptotic analysis; the flexibility and generality of this combination extends balance modeling to a broader range of systems.

Our approach begins with a governing equation, which might be derived from fundamental physics (e.g. Maxwell's equations or the Navier-Stokes equations) but could also result from a model discovery procedure~\cite{Schmidt2009science,Brunton2016pnas,Rudy2017sa}.
These governing equations are physical models capable of describing a wide range of phenomena.
However, it is well understood that the full complexity of such models is not always necessary to describe the local behavior of a system.
In many regimes the dynamics are governed by just a subset of the terms involved in the global description.

We introduce the idea of an equation space, where each coordinate is defined by one of the terms in the governing equation.
Each term may be evaluated individually at any point in space and time, resulting in a vector with each entry corresponding to a term in the governing equation.
We define a dominant balance regime as a region where the evolution equation is approximately satisfied by a subset of the original terms in the equation; the remaining terms may be safely neglected.
When a point in the field is approximately in dominant balance, the equation space representation of the field will have near-zero entries corresponding to negligible terms.
Clearly, the equation space representation of a field is not unique; a fluid flow might be represented by velocity, vorticity, or streamfunction, for example.
The interpretation of the dominant physics therefore depends on the choice of an appropriate governing equation for the application.

Dominant balance physics thus has a natural geometric interpretation in equation space, allowing standard machine learning tools to automatically identify regions where groups of terms have negligible contributions to the local dynamics.
From this perspective, a dominant balance regime is characterized by a cluster of points that have significant covariance in directions of equation space corresponding to active physical processes.
The covariance structure of this cluster is sparse in the sense that there is weak variation in directions that represent the negligible terms.
This corresponds to the mathematical condition that the governing equation is approximately satisfied by a subset of its terms in a local region.

While such dominant balance regimes might be identified by many possible algorithms, we choose to cluster the data using Gaussian mixture models (GMM)~\cite{Bishop2006book} and then extract a sparse approximation to the direction of maximum variance for each cluster using sparse principal components analysis (SPCA)~\cite{Zou2012}.
We take the active terms in each cluster to be those that correspond to nonzero entries in the sparse approximation to the leading principal component.

In simple cases, this two-step GMM-SPCA procedure may be equivalent to applying a hard threshold, where a term is considered active if it exceeds some small value.
However, our approach considers the local, relative importance of terms, whereas thresholding describes global, absolute importance.
This distinction is important in multiscale systems where the scale of the dynamics varies significantly throughout the domain.

The data-driven approach to dominant balance analysis generalizes traditional methods in several critical directions.
First, it does not rely on any explicit assumption of asymptotic scaling.
Second, the clustering method yields pointwise estimates of the spatiotemporally local dominant balance not afforded by traditional scaling analysis in complex geometries.
Third, while many dominant balance regimes have been proposed or assumed based on heuristic or intuitive arguments, this method provides an objective, reproducible approach to testing these hypotheses.
Finally, the probabilistic Gaussian mixture modeling framework is fully compatible with the relative nature of dominant balance analysis, providing natural estimates of uncertainty in the identified balance

~
\newpage
\section{Unsupervised dominant balance identification} \label{sec:method}

In many fields of physics, painstaking analyses have produced models that are capable of describing a wide range of physical phenomena.
However, it is well understood that the full complexity of such models is not always necessary to describe the local behavior of a system.
We find that in many regimes the dynamics are governed by just a subset of the terms involved in the global description.
A general evolution equation for the field $u(x, t) $ on the domain $(x, t) \in \mathcal{D}$ can be written as
\begin{equation} \label{eq:evolution}
\mathcal{N}(u) = \sum_{i=1}^{K} f_i(u, u_x, u_{xx}, \dots, u_t, \dots) = 0.
\end{equation}
For example, the viscous Burgers` equation is
\begin{equation}
\mathcal{N}(u) = u_t + u u_x - \nu u_{xx} = 0
\end{equation}
We represent the equation in this implicit form rather than the typical $u_t = \mathcal{F}(u)$ form for two reasons.
First, it includes arbitrary PDEs which are not easily expressed in the standard form, such as the generalized nonlinear Sch\"odinger equation in Sec.~\ref{sec:results-optics}.
Second, this form highlights the fundamental balance of the equation; all terms must sum to zero.
If some subset is are dominant, the rest must be relatively small.
We assume the total number of terms $K$ is known, either from the physical model or as the result of a model selection procedure~\cite{Mangan2017}.

Consider an ``equation space" where each coordinate is defined by one of the $K$ terms in Eq. \eqref{eq:evolution}.  
At each point $(x,t)$ in space and time, each of the $K$ terms $f_i$ in the governing equations \eqref{eq:evolution} may be evaluated at $u(x,t)$, resulting in a vector $\mathbf{f} \in \R^K$:
\begin{equation} \mathbf{f}(x, t) = \begin{bmatrix}
f_1 (u(x, t), \dots) & f_2(u(x, t), \dots) & \cdots & f_K(u(x, t), \dots)
\end{bmatrix}^T.
\end{equation}
By construction, $ \mathbbm{1}^T \mathbf{f}(x, t) =  \mathcal{N}(u) = 0 $ for all $ (x, t) \in \mathcal{D}$.
Simulated or measured field data is typically discretized, so the domain is approximated by $N$ spacetime points: ${
	\mathcal{D} \approx \left\{ (x, t)^j \hspace{0.2cm} |\hspace{0.2cm} j = 1, 2, \dots, N \right\}}
$.
The field at each of these points corresponds to a point in equation space. 

We define a dominant balance regime as a region $ \mathcal{R} \subset \mathcal{D}$ where the evolution equation is approximately satisfied by a subset  of $p < K$ of the original terms in the equation; the remaining terms may be neglected.
In this case $\mathbf{f}(x, t)$ will have near-zero entries corresponding to negligible terms when $(x, t) \in \mathcal{R}$.
Geometrically, the field is approximately restricted to $p$ of the original $K$ dimensions of the equation space, resulting in a subspace that is aligned with the active $p$ terms.

This geometric perspective on dominant balance physics leads naturally to segmentation via unsupervised clustering.
For example, the Gaussian mixture model (GMM) framework learns a probabilistic model by assuming the data are generated from a mixture of Gaussian distributions with different means and covariances \cite{Bishop2006book}.
The learned covariances for each cluster can then be interpreted in terms of active and inactive terms in the evolution equation.  The $N$ spacetime points in $\mathcal{D}$ are used to train a mixture model; the algorithm treats points from a dominant balance regime as if they were generated from a distribution with near-zero variance in the directions corresponding to negligible terms.
Data beyond the original inputs can efficiently be assigned to a balance model using the trained GMM.

In practice, there is no reason to expect the points will even approximate a mixture of Gaussian distributions.  We therefore expect that the number of clusters required to capture all of the relevant physics will exceed the number of distinct balance regimes, resulting in redundant clusters.  Furthermore, there is some ambiguity in the interpretation of ``near-zero variance".
We address both of these issues using sparse principal components analysis (SPCA) \cite{Zou2012}, which uses $\ell_1$ regularization to extract a sparse approximation to the leading principal component.
If a cluster describes a dominant balance regime, it should be well-described by its direction of maximum variance.
Moreover, this leading principal component should have many near-zero entries.
We apply SPCA to the set of points in each GMM cluster and take the active terms in the cluster to be those which correspond to nonzero entries in the sparse approximation to the leading principal component.
The number of models can then be reduced by grouping clusters with the same set of active terms (or equivalently, the same sparsity pattern in the SPCA approximation).

Dominant balance identification can be seen as a localized active subspace analysis in equation space \cite{Constantine2014}.
Rather than assuming that there is a global decomposition into approximately active and inactive subspaces, we simultaneously search for subspaces corresponding to different balance relations and the regions of the domain where the dynamics are well-described by this subspace.

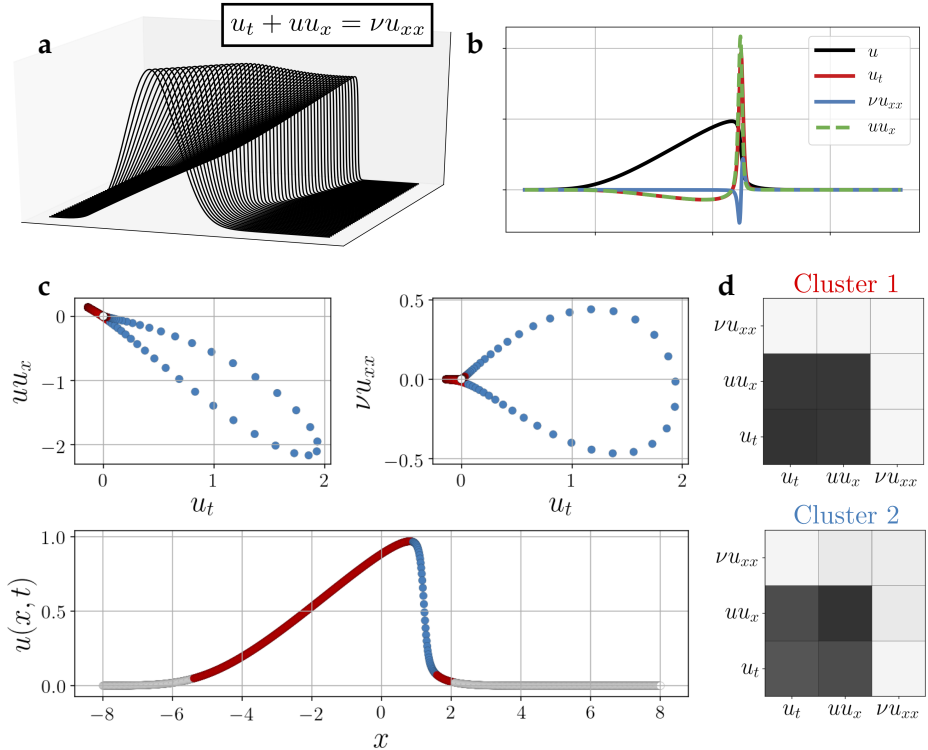
\begin{figure}
	\centering
	\begin{overpic}[width=0.75\linewidth]{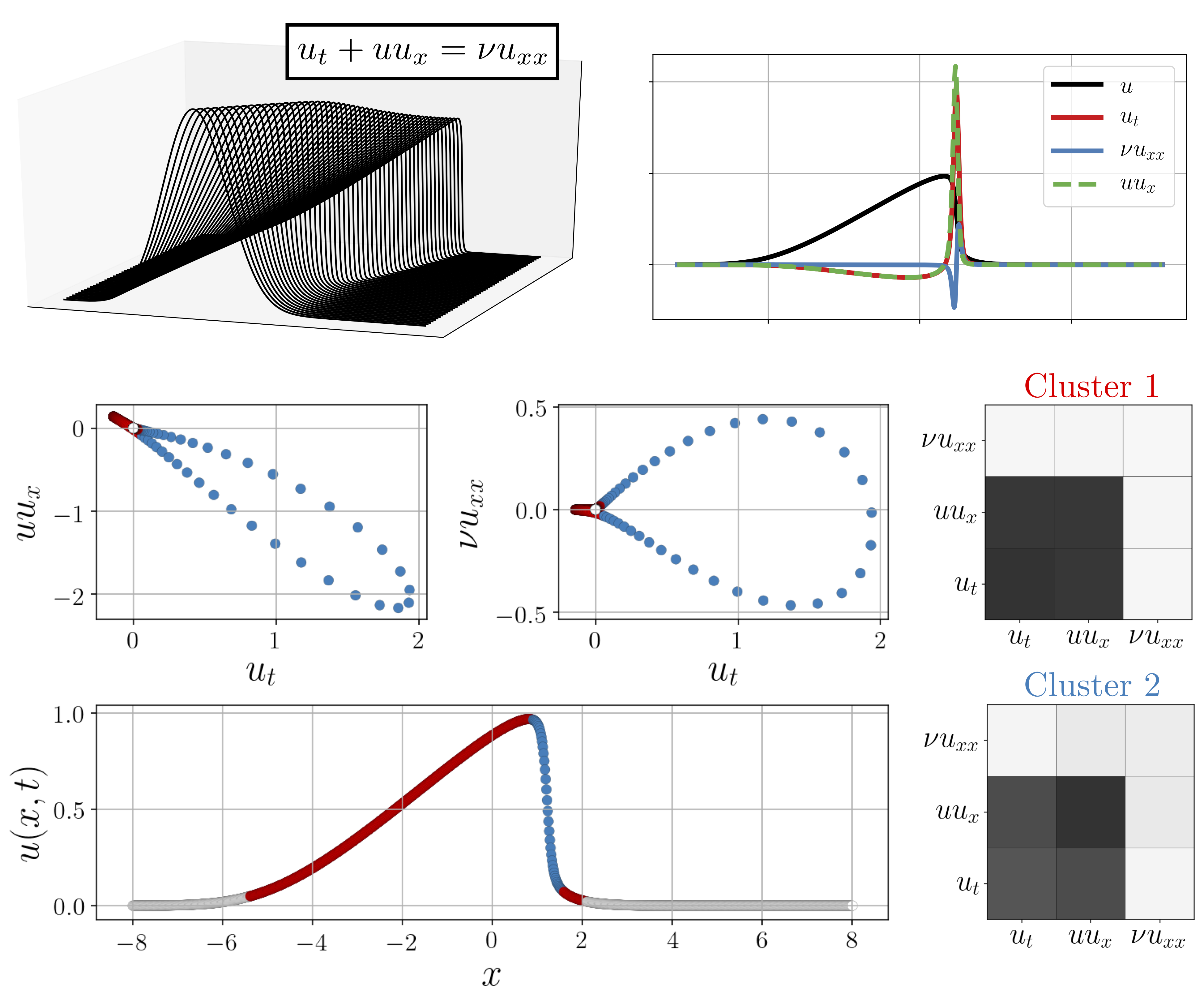}
		\put(4, 76){\textbf{a}}
		\put(50, 76){\textbf{b}}
		\put(4, 50){\textbf{c}}
		\put(77, 50){\textbf{d}}
	\end{overpic}
	\vspace{-.1in}
	\caption{
		Example of dominant balance identification on the viscous Burgers' equation (a), with constituent terms shown in (b). The viscous term acts to diffuse sharp gradients and prevent formation of a discontinuous shock, but away from the shock front the dynamics are essentially inviscid.
		Away from the shock front, the field is approximately restricted to the $\nu u_{xx} = 0$ plane (c).
		This is reflected in the covariance matrices learned by the Gaussian mixture model (d).
	}
	\label{fig:burgers}
\end{figure}

\subsection*{Burgers' equation}
For example, one of the simplest models that demonstrates dominant balance is the viscous Burgers' equation, shown in Fig. \ref{fig:burgers}.
Shocks form from the nonlinear advection and are dissipated by the viscous term.
Away from the shock front, however, the gradients of the field are relatively weak, so viscosity does not contribute significantly to the dynamics.
Figure \ref{fig:burgers} demonstrates the balance identification procedure applied to a snapshot of the viscous Burgers' equation example.
Most of the field is classified into two clusters, corresponding to either no dynamics or an inviscid balance between acceleration and advection.
Only a narrow slice along the shock front belongs to a cluster in which viscosity is active.

In simple cases, this two-step GMM-SPCA procedure might be replaced with a hard threshold; if a term exceeds some value $\epsilon$ it is ``on".
However, the proposed method offers two main advantages over thresholding.
First, the idea of dominant balance has a natural geometric interpretation in equation space, thereby avoiding setting an arbitrary threshold for which diagnostics and interpretation may not be straightforward.
Second, our method considers the \textit{local, relative} importance of terms, whereas thresholding describes \textit{global, absolute} importance.
For example, this distinction is significant in multiscale systems with some background process underlying intermittent bursts of activity.
The intermittency is dominated by a balance between terms which may be much larger than the background process, although the dynamics during quiescent periods would be determined primarily by the background process.
In this case an absolute thresholding method would either choose the background process to be always on or always off, whereas a relative approach recognizes that the dominant local balance simply changes during the intermittent activity.
This is illustrated in Sec. \ref{sec:results-neuron}, where we investigate a Hodgkin-Huxley-type model of spiking neuron, generalized to introduce multiscale bursting behavior.

\section{Results}
\label{sec:results}

\begin{figure}
	\centering
	\begin{overpic}[width=\linewidth]{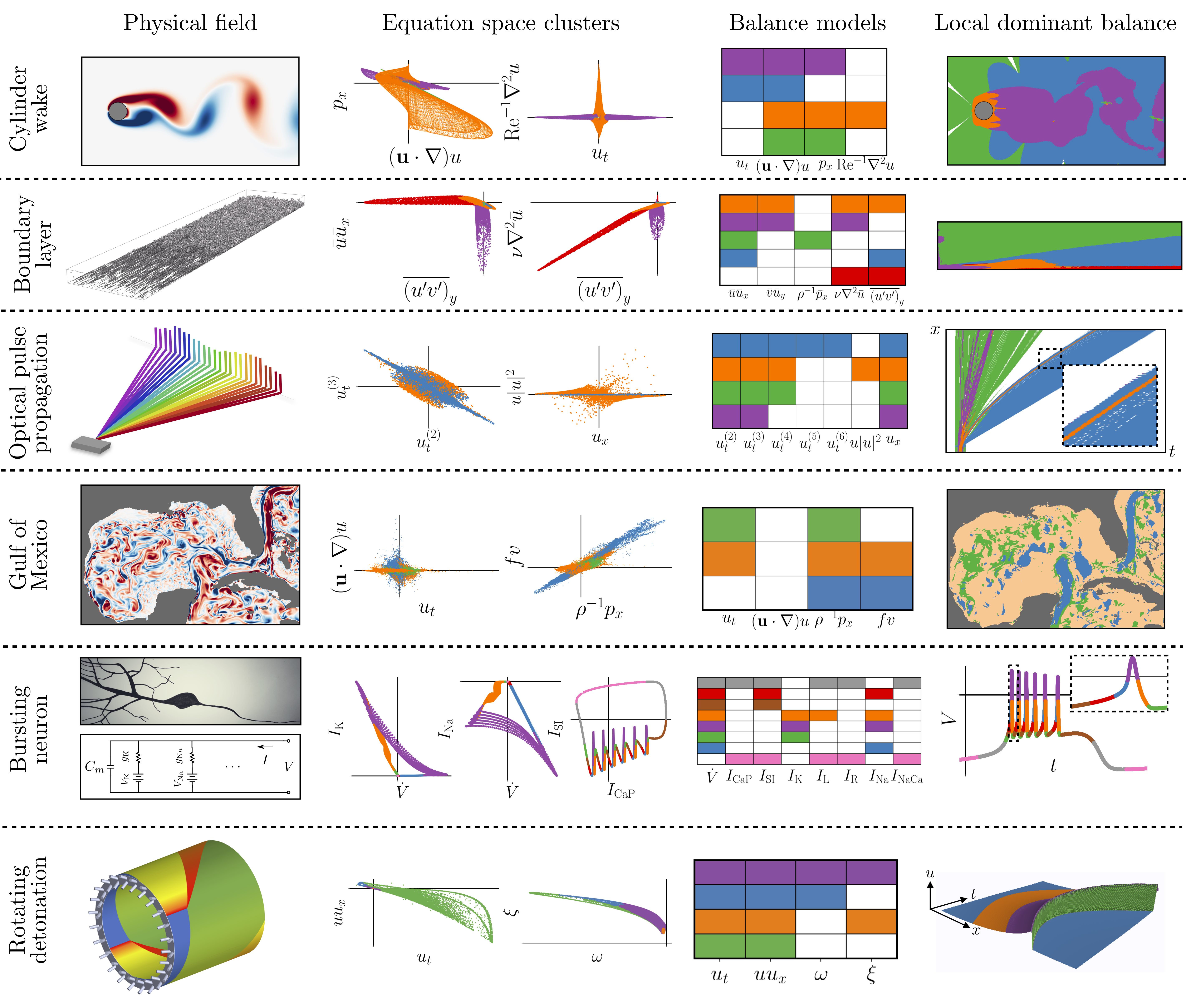}
	\end{overpic}
	\vspace{-.3in}
	\caption{Dominant balance physics identified across a range of systems.
	For each case, a visualization of the system is shown on the left, followed by 2D views of the feature space colored by the identified balance relation, a key describing the active terms in each model, and the original field colored by the local balance.
	From top: a bluff body wake at moderate Reynolds number, a boundary layer in transition to turbulence, pulse propagation in an optical fiber, surface currents in the Gulf of Mexico, and a Hodgkins-Huxley model for an intrinsically bursting neuron.
	}
	\label{fig:summary}
\end{figure}

We now apply the dominant balance identification method to a range of physics with varying complexity: unsteady vortex shedding past a cylinder at Reynolds number 100; the mean field of a turbulent boundary layer; optical pulse propagation in supercontinuum generation; geostrophy in the Gulf of Mexico; and a Hodgkin-Huxley-type model of a biological neuron.
Figure \ref{fig:summary} shows a summary of the results, including slices of the equation space representations, identified balance models, and segmented fields.
In each case, the results are consistent with classical scaling analyses and known physical behavior. 
Descriptions of the models and code used to generate this data are presented in Appendix A and are available online.  

\subsection{Flow past a circular cylinder at $\Re=100$}

\paragraph{Governing equations and analytic scaling}  Flow past a cylinder at moderate Reynolds number is a prototypical flow configuration for bluff body wakes.  The wake transitions from steady laminar flow to periodic vortex shedding via a Hopf bifurcation at $\Re \approx 47$. %
The transition from linear instability to a stable limit cycle is itself a fascinating example of dominant balance in fluid mechanics and dynamical systems.
The quadratic nonlinearity, initially inactive in the linear regime, mediates energy transfer between the mean flow and instability modes, deforming both until an energy balance is reached in the periodic limit cycle.
This nonlinear stability mechanism was first described by Stuart and Landau \cite{Stuart1958jfm, Landau1959book} and later employed for reduced-order modeling \cite{Noack2003jfm}.

Even in the stable limit cycle, however, the local dynamics of the flow vary widely throughout the domain, highlighting mechanisms that give rise to von K\`{a}rm\`{a}n-type vortex streets in a wide variety of flows.
This unsteady, incompressible, viscous flow is governed by the two-dimensional Navier-Stokes equations:
\begin{equation}
\label{eq:2d-ns-dimensions}
\mathbf{\tilde{u}}_t + (\mathbf{\tilde{u}} \cdot \nabla) \mathbf{\tilde{u}} = -\frac{1}{\rho} \nabla \tilde{p} + \nu \nabla^2 \mathbf{\tilde{u}},
\end{equation}
where $\mathbf{\tilde{u}}$ is the velocity field, $ \tilde{p} $ is the pressure, $\rho$ is the density, and $\nu$ is kinematic viscosity.
Of course, these equations themselves involve some degree of approximation, ignoring effects such as compressibility and gravity, making use of the Newtonian form of the stress tensor, and assuming Fickian diffusion, though they have proven highly accurate when applied in the correct regime.
Nevertheless, there are distinct regimes in this simple wake flow.

For the wake behind a circular cylinder, the most relevant scales are the cylinder diameter $L$ and free-stream velocity $U$.
Dimensional analysis then suggests that
\begin{equation*}
\mathbf{\tilde{u}} \sim U, \hspace{1cm} \tilde{p} \sim \nu U^2, \hspace{1cm} \nabla \left( \cdot \right)  \sim \frac{1}{L}, \hspace{1cm} \pdv{}{t}  \left( \cdot \right) \sim \frac{U}{L}.
\end{equation*}
Nondimensionalizing with respect to these scales, we find that the viscous term is smaller than the others by a factor of the Reynolds number, $\Re = UL/\nu$,  resulting in the familiar nondimensional form of the Navier-Stokes equations:
\begin{equation}
\label{eq:2d-ns}
\mathbf{u}_{t} + (\mathbf{u} \cdot \nabla) \mathbf{u} = -\nabla p + \frac{1}{\Re} \nabla^2 \mathbf{u}.
\end{equation}
The variables and operators have been nondimensionalized according to the previous scales.
For even moderately large Reynolds numbers, we would expect the flow to behave in an approximately inviscid manner away from the cylinder.
Thus, structures formed in the near-wake region will be advected downstream by the mean flow with only weak dissipation, as observed in the vortex street.

Near the cylinder, the no-slip boundary conditions due to viscosity change the behavior qualitatively.  If we examine the flow at a point a distance $\delta \ll L $ from the wall, then $\delta$ is a more appropriate length scale for the gradients.  However, since the near-wall flow varies on a similar timescale to the wake, suppose that $U/L$ is still a good scale for the time derivative.
The various terms then scale as
\begin{equation*}
\mathbf{\tilde{u}}_t \sim \frac{U^2}{L}, \hspace{1cm} (\mathbf{\tilde{u}} \cdot \nabla) \mathbf{\tilde{u}} \sim \frac{U^2}{\delta} \hspace{1cm} -\frac{1}{\rho} \nabla \tilde{p}  \sim \frac{U^2}{\delta}, \hspace{1cm} \nu \nabla^2 \mathbf{\tilde{u}} \sim \frac{\nu U}{\delta^2}.
\end{equation*}
We find that the acceleration term is now smaller by a factor of $\delta / L $, and expect the viscous term to be balanced by advection and the pressure gradient. 
The relatively strong gradients near the wall give rise to the vortex structures which characterize the wake.

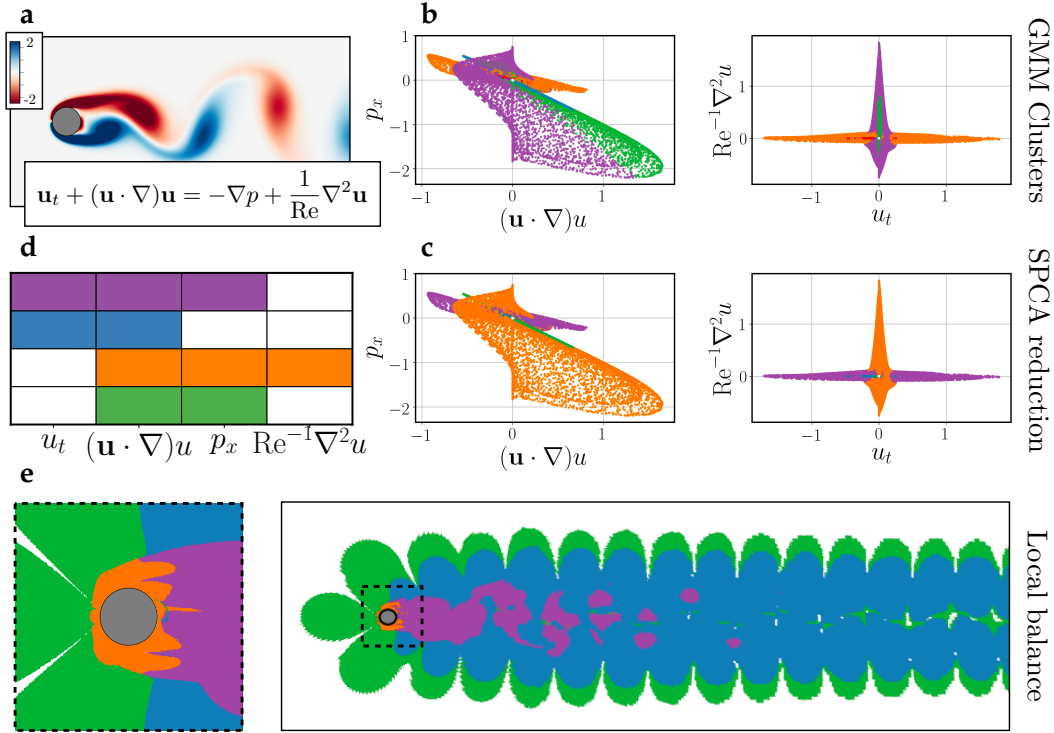
\begin{figure}
	\begin{center}
		\begin{overpic}[width=0.85\linewidth]{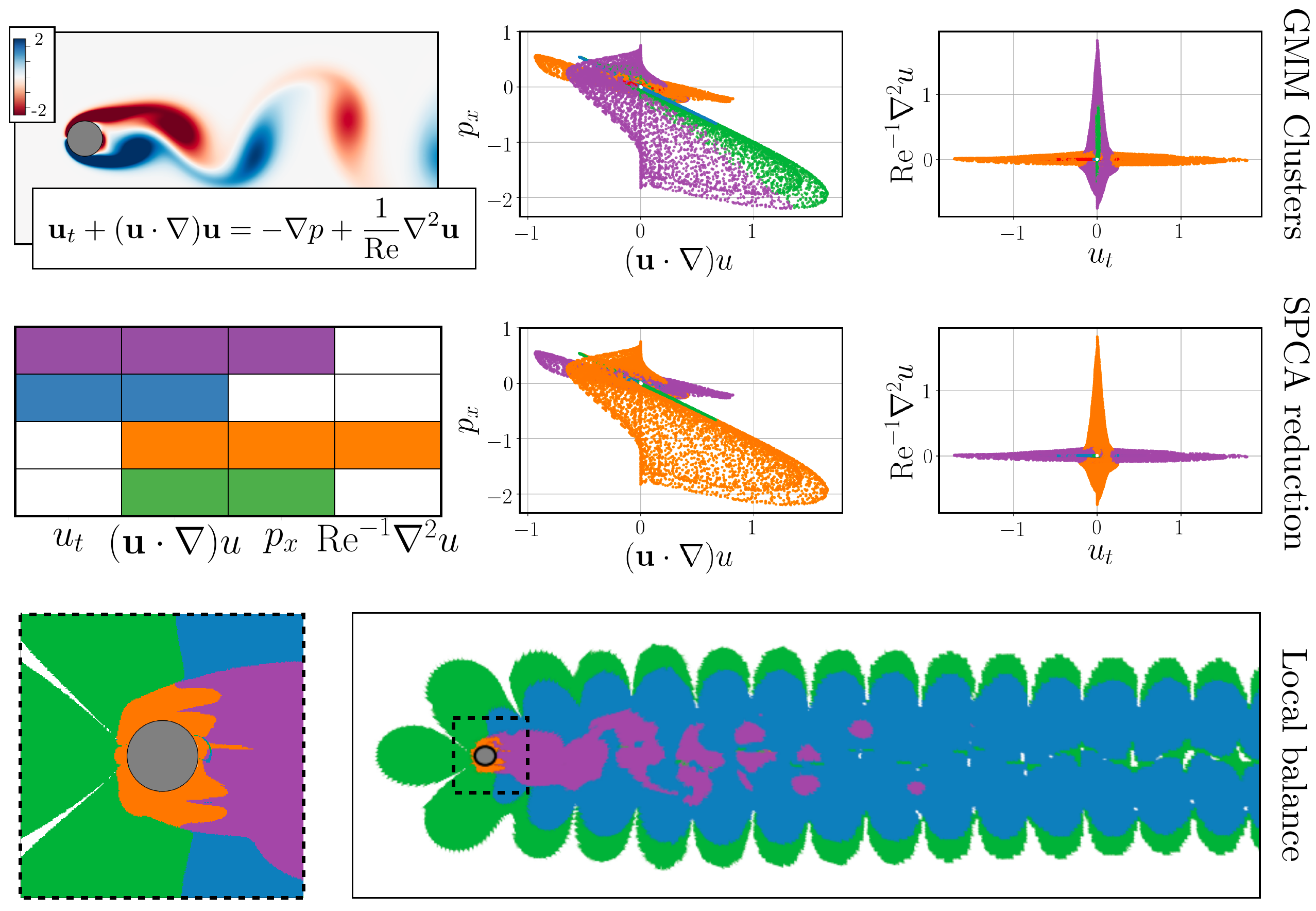}
			\put(2, 68){\textbf{a}}
			\put(40, 68){\textbf{b}}
			\put(2, 46){\textbf{d}}
			\put(40, 46){\textbf{c}}
			\put(2, 24.5){\textbf{e}}
		\end{overpic}
		\vspace{-.2in}
	\end{center}
	\caption{Vorticity snapshot for the wake behind a cylinder at $\Re = 100$ (a).
		A Gaussian mixture model (GMM) assigns field points to clusters by looking for groups with distinct mean and covariance (b).
		For instance, some clusters vary mainly in the acceleration-advection directions, while others vary principally in the viscous-advection directions.
		We would expect these to represent the far-field and boundary regions, respectively.
		This is confirmed by the sparse principal components analysis (SPCA) reduction, where clusters with significant nonzero variance in the same directions are grouped together (c).
		These directions can be interpreted as active terms in the balance relation (d).
		As anticipated, the region near the cylinder is dominated by a balance between viscosity and advection and pressure forces, while the far wake is approximately inviscid (e). }
	\label{fig:cyl100}
\end{figure}

\paragraph{Identified dominant balance}
Figure \ref{fig:cyl100} shows an example vorticity field along with views of the 4D equation space corresponding to Eq. \eqref{eq:2d-ns}.
Although the method treats space and time equivalently, here we freeze time and explore a single snapshot; since the flow is periodic we expect the results to be representative.
The visualization in equation space clearly reveals signatures of balance relations.
One set of GMM clusters is nearly restricted to the the zero-viscosity plane, while another has reduced variance in the acceleration direction.  
The sparse approximations to the leading principal components of each cluster confirms this intuition; we use SPCA to construct balance models by grouping the Gaussian models with non-negligible variance in the same directions.
As expected, the far wake is approximately inviscid, while the region near the cylinder is dominated by a balance between viscosity, pressure, and advection.
This method also identifies other approximate regions, such as a low-pressure-gradient balance between acceleration and advection (blue), slowly varying potential flow (green), and a far-field region with near-zero dynamics (white).

\paragraph{Nonlinear stability}
The cylinder wake at moderate Reynolds number is of particular interest in the reduced-order modeling community because it is a canonical example of a self-limiting instability exhibiting the Stuart-Landau nonlinear stability mechanism~\cite{Noack2003jfm, Sipp2007jfm, ManticLugo2014prl}.
The steady-state solution $(\mathbf{U}, P)$, defined as\footnote{All fields are understood to be divergence-free; here we give only the momentum equations for brevity.}
\begin{equation}
(\mathbf{U} \cdot \nabla ) \mathbf{U} = - \nabla P + \Re^{-1} \nabla^2 \mathbf{U},
\end{equation}
becomes unstable to infinitesimal perturbations at $\Re_c \approx 46$.
At this critical parameter value, the flow undergoes a Hopf bifurcation~\cite{Jackson1987, Provansal1987}; the asymptotic solution is the limit-cycle vortex shedding explored above.

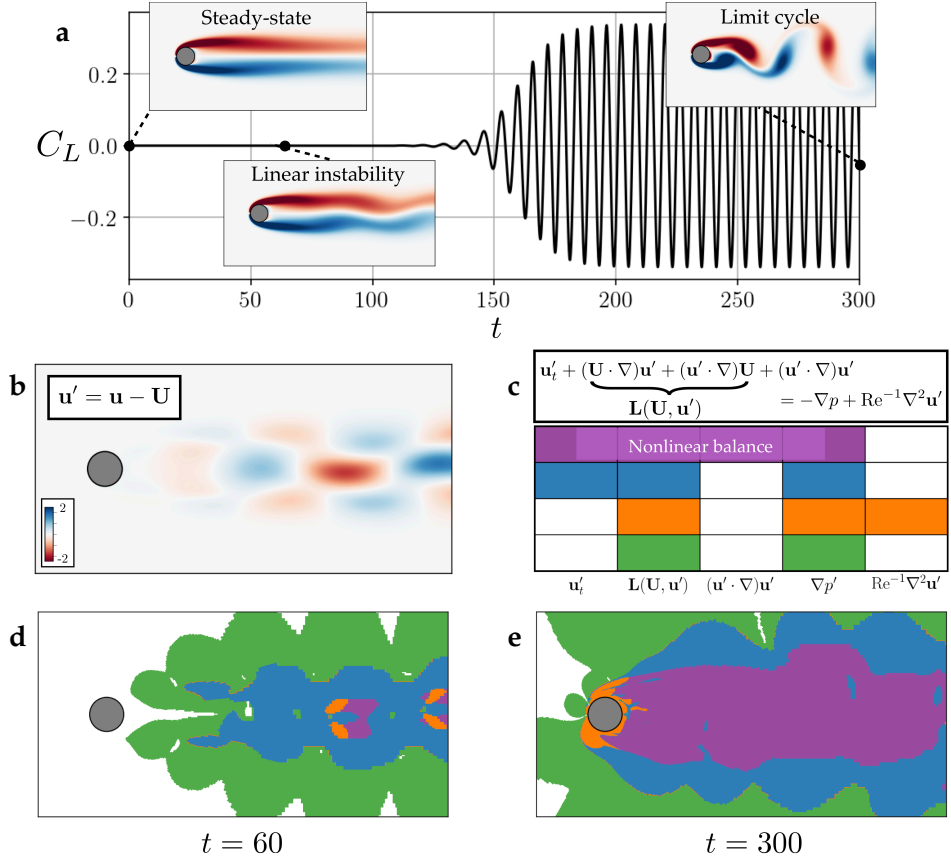
\begin{figure}
	\begin{center}
		\begin{overpic}[width=0.75\linewidth]{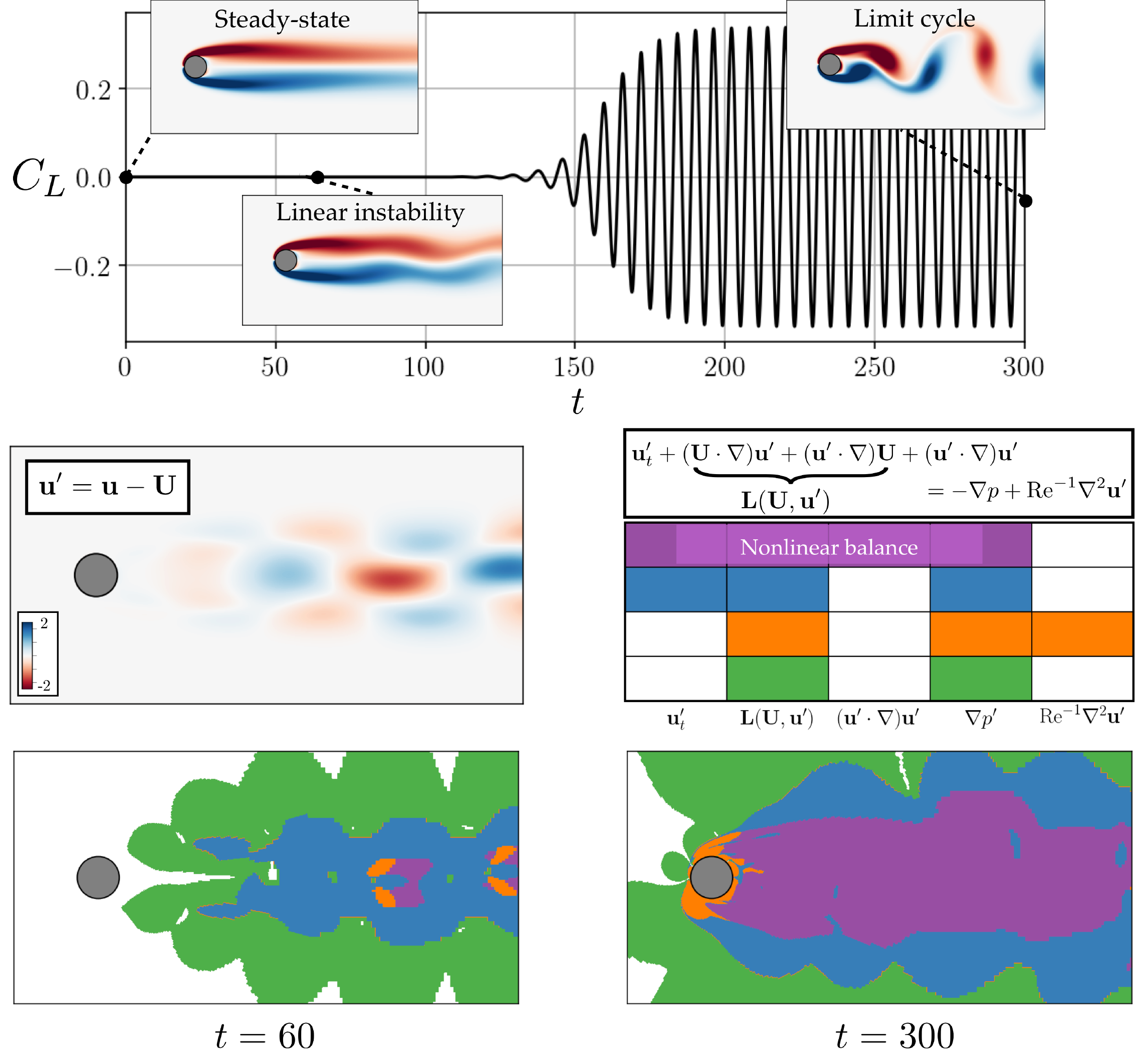}
			\put(3, 88){\textbf{a}}
			\put(-2, 50){\textbf{b}}
			\put(51.5, 50){\textbf{c}}
			\put(-2, 23){\textbf{d}}
			\put(51.5, 23){\textbf{e}}
		\end{overpic}
		\vspace{-.2in}
	\end{center}
	\caption{
		The role of nonlinearity in stabilizing a von K\`arm\`an vortex street. 
		The transient flow evolves from an unstable steady-state through the exponential growth of a linear instability mode to the vortex shedding limit cycle (a, visualized with lift coefficient $C_L$).
		Only one balance is identified that includes the nonlinear term $( \mathbf{u}' \cdot \nabla ) \mathbf{u}'$.
		This balance does not appear significantly in the linear growth regime (d), consistent with the correspondence between the fluctuations and the instability mode.
		The fully saturated limit cycle is dominated by this balance, however, confirming the interpretation of the stabilizing feedback loop mediated by the nonlinearity.
	}
	\label{fig:cyl-transient}
\end{figure}

The mechanism by which the exponential energetic growth of the linear instability transitions to a stable limit cycle may be understood in terms of the Stuart-Landau mean field theory.
An arbitrary flow field may be decomposed into the steady-state ``base" flow $\mathbf{U}$ and a time-varying perturbation $\mathbf{u}'$:
\begin{equation}
\mathbf{u}(\mathbf{x}, t) = \mathbf{U}(\mathbf{x}) + \mathbf{u}'(\mathbf{x}, t).
\end{equation}
A similar expansion can be applied to the pressure field.
Expanding the momentum equations, we find that the fluctuations evolve according to
\begin{equation}
\label{eq:NS-perturbation}
\mathbf{u}'_t + (\mathbf{U} \cdot \nabla ) \mathbf{u}' + (\mathbf{u}' \cdot \nabla) \mathbf{U} + (\mathbf{u}' \cdot \nabla )\mathbf{u}' = -\nabla p' + \Re^{-1} \nabla^2 \mathbf{u}'.
\end{equation}
The only part of these equations that is nonlinear in $\mathbf{u}'$ is the advection term $(\mathbf{u}' \cdot \nabla) \mathbf{u}'$.
Linear stability analysis proceeds by assuming the fluctuations are weak enough that this term is negligible.

However, the exponential growth of unstable modes implies that eventually this assumption cannot hold; in order for the energy of the flow to be bounded at long times the nonlinearity must play a stabilizing role.
This is typically conceptualized as a mean field deformation.
The mean of the limit cycle has a much shorter recirculation region than the steady-state solution and is approximately neutrally stable~\cite{Barkley2006epl}.
The role of the nonlinearity is thus understood as a feedback mechanism.
The base flow is deformed in a manner that reduces the growth rate of the instability.
The two come into energetic balance on the limit cycle, where the mean flow becomes neutrally stable; this is the basis for the ``self-consistent" mean field modeling approach~\cite{ManticLugo2014prl}.
Although this argument is based on a linear perturbation analysis of a smoothly deforming base flow, it can be confirmed more rigorously close to the critical Reynolds number with a weakly nonlinear expansion~\cite{Sipp2007jfm}.

The role of the nonlinearity can also be examined from the perspective of dominant balance, without assuming either linearity or proximity to the bifurcation.
Figure~\ref{fig:cyl-transient} shows the transient evolution of the cylinder wake, from the unstable steady-state through the linear growth region and ultimately to the vortex shedding limit cycle.
We apply dominant balance anlysis to the evolution equation for the base-subtracted snapshots, Eq.~\eqref{eq:NS-perturbation} in both the linear instability regime and the post-transient limit cycle.
The method identifies four dominant balances: a far-field steady balance between base flow advection and the pressure gradient (green), a steady viscous region (orange), an inviscid linear balance (blue), and finally a balance including the nonlinear term (purple).

The nonlinear balance is largely absent from the transient growth snapshot.
This is expected from the assumptions of linear stability analysis; the fluctuations at this stage are nearly identical to the linear instability mode~\cite{Barkley2006epl}.
On the other hand, the wake region of the post-transient snapshot is largely nonlinear, which confirms the picture of the role of nonlinearity in stabilizing the limit cycle.

Moreover, this analysis clearly demonstrates the instantaneous local balance of the terms, without relying on either assumptions of linearity or limiting parameter regimes.
This distinguishes it from analyses such as weakly nonlinear expansions, self-consistent modeling, and energy balance arguments.
Furthermore, this method is non-intrusive and could be applied to either experimental or numerical observations with more complex decomposition structures such as the harmonic balance expansion approach to nonlinear resolvent analysis~\cite{Rigas2020jfm}.

\paragraph{Spurious terms}
A key feature of the equation space representation of the evolution equation~\eqref{eq:evolution} is the constraint that all coordinates must sum to zero, provided the terms are computed correctly and the correct governing equation is chosen.
This implies a \textit{linear} covariance structure, even when the dynamics are strongly nonlinear; each term must be balanced by a linear combination of the others.

This constraint is not explicitly enforced in the present method, although it is the reason that the search for sparse linear subspaces with the GMM/SPCA algorithm is a natural approach.
However, this observation is one avenue by which discrepancies in the governing equation may be diagnosed.
If one term cannot be balanced by a linear combination of the others, it is either a spurious term or the governing equation is incomplete.

\begin{figure}
	\begin{center}
		\begin{overpic}[width=0.85\linewidth]{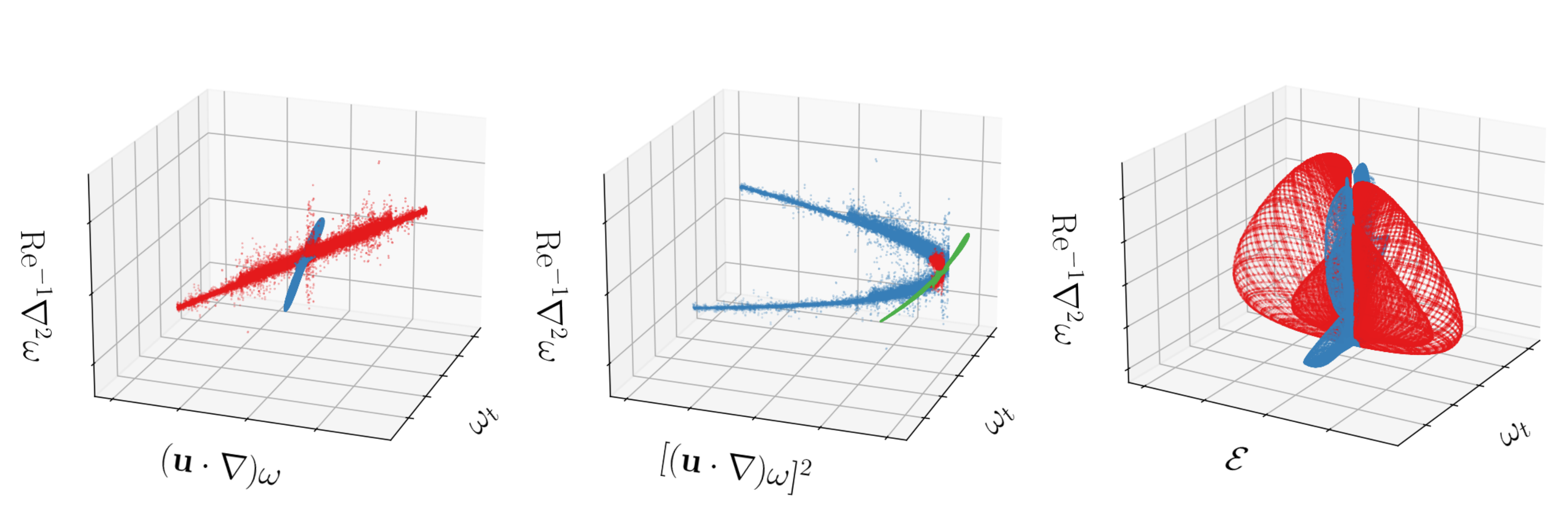}
			\put(5, 30){\textbf{a}\hspace{0.5cm}Correct terms}
			\put(36, 30){\textbf{b}\hspace{0.2cm}Incorrect: quadratic}
			\put(71, 30){\textbf{c}\hspace{0.2cm}Incorrect: energy}
		\end{overpic}
		\vspace{-.2in}
	\end{center}
	\caption{ 
		Effect of including incorrect terms in the governing equations.
		Even with strongly nonlinear dynamics, the fields vary along sparse linear subspaces in the equation space representation (a).
		By definition, the equation must be balanced by a linear combination of the terms.
		Introducing spurious terms such as the square of the convective term (b) or the kinetic energy (c) leads to significant departure from this behavior and may be readily diagnosed.
		In these cases there is either a clear nonlinear covariance structure, or the equation fails to close.
	}
	\label{fig:spurious-terms}
\end{figure}

For example, in two dimensions the Navier-Stokes equations can be reduced to the form:

\begin{equation}
\omega_t + (\mathbf{u} \cdot \nabla) \omega = \frac{1}{\Re} \nabla^2 \omega,
\end{equation}
where $\omega = \nabla \times \mathbf{u}$ is the normal vorticity.
As shown in Fig.~\ref{fig:spurious-terms}a, the three-dimensional equation space representation has a clear linear covariance structure.
One cluster (blue) has negligible viscosity and varies primarily in the unsteady-advection directions, while the other (red) is dominated by viscosity and advection, with only weak time variation.

However, if the advection term $(\mathbf{u} \cdot \nabla) \omega$ is replaced by either its square or by the total kinetic energy $\mathcal{E} = \frac{1}{2} ( \mathbf{u} \cdot \mathbf{u})$, the fundamental requirement that Eq.~\eqref{eq:evolution} sums to zero is clearly violated, as shown by Fig.~\ref{fig:spurious-terms}b-c.
The clustering procedure does not give physically meaningful results, and this is easily diagnosed by considering the linear closure constraint.

\subsection{Turbulent boundary layer}
\label{sec:results-boundary}
One of the major breakthroughs in the study of fluid mechanics in the 20th century was the development of boundary layer theory \cite{Schlichting1955, Pope2000book}.  In many practical applications fluids can be treated as inviscid, but close to solid boundaries strong velocity gradients lead to significant viscous forces.
Prandtl showed in 1904 that careful scaling analysis applied to the governing Navier-Stokes equations reveals distinct regimes where the behavior of the fluid is essentially determined by a small subset of the full equations.
In turn, these balance relations can be used to derive powerful scaling laws such as the so-called ``law of the wall".

Although such analyses can be intractable for general turbulent flows, one of the most important canonical configurations is zero pressure gradient flow over a flat plate parallel to the free stream velocity.  
The zero pressure gradient ensures that the free-stream velocity is constant in the streamwise direction at large distances from the wall.
This flow is statistically two-dimensional; the configuration does not vary in the cross-stream direction so the mean flow only varies in the streamwise and wall-normal directions.

\paragraph{Governing equations and analytic scaling}
After performing the Reynolds decomposition of the variables into mean and fluctuating components, e.g. $ u = \bar{u} + u' $, the mean flow is determined by the Reynolds-averaged Navier-Stokes (RANS) equations.  For the streamwise mean velocity $ \bar{u}$, the equation is
\begin{equation} \label{eq:rans}
\bar{u} \pdv{\bar{u}}{x} + \bar{v} \pdv{\bar{u}}{y} = \rho^{-1} \pdv{\bar{p}}{x} + \nu \nabla^2 \bar{u}  - \pdv{y}\overline{u' v'} - \pdv{x}\overline{u'^2}.
\end{equation}
The terms on the left represent mean flow advection, while those on the right are the pressure gradient, viscosity, wall-normal Reynolds stress, and streamwise Reynolds stress, respectively.

One of the challenges in studying this flow is that there are multiple length scales.
Following \cite{Holmes1996}, we may consider a streamwise length scale $L$, a wall-normal length scale $\ell$, and a viscous length scale $\eta = \nu / u_\tau$, where $u_\tau$ is the ``friction velocity" associated with the shear stress at the wall.

Beginning with the ``outer" region of the boundary layer (where $y \gg \eta$), suppose the mean streamwise velocity $\bar{u}$ scales with the free stream $U_\infty$, while the turbulent fluctuations $u', v'$ scale with $u_\tau$.
As with the previous example, assume that the derivatives scale with the corresponding length scale, so that for instance $(\cdot)_y \sim 1 / \ell $.
For instance, the continuity equation $\bar{u}_x + \bar{v}_y = 0 $ implies that $\bar{v} \sim U_\infty (\ell / L) $.
By this reasoning typically we would expect the mean velocity gradient $\bar{u}_y$ to scale with $U_\infty/\ell$, but as argued in \cite{Holmes1996}, the gradients in the outer part of the layer are much weaker than near the wall, and empirically a better estimate is $\bar{u}_y \sim u_\tau/\ell$.
Then for the streamwise momentum equation we find 
\begin{equation*}
\bar{u} \bar{u}_x \sim \frac{U_\infty^2}{L},
\hspace{0.5cm}
\bar{v} \bar{u}_y \sim \frac{u_\tau U_\infty}{L}
\hspace{0.5cm}
\nu \bar{u}_{xx} \sim  \frac{\nu U_\infty}{L^2},
\hspace{0.5cm}
\nu \bar{u}_{yy} \sim  \frac{\nu u_\tau}{\ell^2},
\hspace{0.5cm}
(\overline{u' v'})_y \sim \frac{u_\tau^2}{\ell},
\hspace{0.5cm}
(\overline{{u'}^2})_x \sim \frac{u_\tau^2}{L},
\end{equation*}
and the pressure gradient is negligible by construction.
Since $L \gg \ell$ we neglect the streamwise Reynolds stress compared to the wall-normal term.
On the other hand, since $U_\infty \gg u_\tau$, we can assume the mean flow advection is dominated by the streamwise component $\bar{u}\bar{u}_x$.
Finally, the viscous terms are smaller than the advection by a factor on the order of the Reynolds number $\Re_L = U_\infty L / \nu \gg 1$.
The outer part of the boundary layer is then determined by an inertial balance between streamwise mean flow advection and wall-normal Reynolds stress:
\begin{equation}
\label{eq:inertial-sublayer}
(\overline{u' v'})_y = - \bar{u} \bar{u}_x.
\end{equation}

However, this relation cannot describe the near-wall regime, where viscosity is known to be important.
In this region we expect the wall-normal derivatives to scale with $(\cdot)_y \sim 1/\eta = u_\tau / \nu$.
As a consequence of the no-slip boundary conditions, in this region the free-stream velocity is not an appropriate scale for the streamwise component and we should instead use the friction velocity $u_\tau$, so that
\begin{equation*}
\bar{u} \bar{u}_x, \bar{v} \bar{u}_y \sim \frac{u_\tau^2}{L},
\hspace{0.5cm}
\nu \bar{u}_{xx} \sim  \left(\frac{\eta}{L}\right) \frac{u_\tau^2}{L},
\hspace{0.5cm}
\nu \bar{u}_{yy} \sim  \left(\frac{L}{\eta}\right) \frac{u_\tau^2}{L},
\hspace{0.5cm}
(\overline{u' v'})_y \sim \left(\frac{L}{\eta}\right) \frac{u_\tau^2}{L},
\hspace{0.5cm}
(\overline{{u'}^2})_x \sim \frac{u_\tau^2}{L}.
\end{equation*}

In this case the wall-normal Reynolds stress is larger than the mean flow advection by a factor of $L / \eta \gg 1$ and must instead be balanced by the viscosity.
Therefore, in a thin viscous sublayer near the wall the dominant balance is
\begin{equation}
\label{eq:viscous-sublayer}
(\overline{u' v'})_y = \nu \bar{u}_{yy}.
\end{equation}

The overall picture is then that the Reynolds stress must be balanced by mean flow advection in the inertial sublayer and by viscosity in the near-wall region.
Outside of the turbulent boundary layer the Reynolds stresses and mean wall-normal velocity are negligible, so small variations, for instance due to incompletely converged statistics, should be described by the balance $  \bar{u} \bar{u}_x = -\rho^{-1} \bar{p}_x $.
In a true zero pressure gradient flow both of these would be zero in the free stream.

\paragraph{Identified dominant balance}

We investigate the dominant balance physics of transitional boundary layer data from a direct numerical simulation~\cite{Zaki2013, Lee2018, Wu2019prf}, openly available from the Johns Hopkins Turbulence Database \cite{Perlman2007jhtdb, Li2008jhtdb}\footnote{https://doi.org/10.7281/T17S7KX8}.
Figure \ref{fig:boundary-layer} shows the equation space clusters and associated dominant balance models for the mean fields.
As with the cylinder example, some sets of points have significantly reduced variance in certain directions of equation space, a strong signature of the dominant balance phenomenon.

The method identifies regions corresponding to the viscous sublayer \eqref{eq:viscous-sublayer}, inertial sublayer \eqref{eq:inertial-sublayer}, and slightly perturbed free stream.
It also identifies a region near the inlet characterized by a lack of Reynolds stresses, suggesting the mean profile here should be consistent with the laminar solution.
The boundaries between balance regimes need not be sharp, however, especially in a transitional flow.
In this case a cluster containing all of the active terms in the zero-pressure-gradient flat plate turbulent boundary layer equation is identified between the laminar inflow region and fully developed turbulence downstream.

Equations \eqref{eq:inertial-sublayer} and \eqref{eq:viscous-sublayer} are a starting point for many of the results of boundary layer theory; from these a range of useful laws can be derived, such as the logarithmic mean velocity profile in the inertial sublayer.
Although we ultimately hope that data-driven balance identification will open new avenues of analysis, we can also use established results to examine the validity of the proposed method.

For example, the dominant length scale $\ell$ in the inertial sublayer is expected to depend on the streamwise coordinate $x$ via a power law $\ell \sim x^{4/5}$ \cite{Schlichting1955}.
It is not usually obvious how to extract a specific value of $\ell$ for which this scaling can be checked.
However, as a rough proxy we may consider the wall-normal coordinate at which the dominant balance changes from that of the inertial sublayer to the free-stream.
Figure \ref{fig:boundary-layer} shows the growth of the inertial sublayer thickness according to this definition along with a power law fit with exponent 0.81, showing close agreement with the expected value of 4/5.
Although this evidence is somewhat circumstantial, it is at least suggestive that the balance model identification procedure reflects the underlying physics.

\paragraph{Self-similarity}

\begin{figure}
	\centering
	\begin{overpic}[width=0.8\linewidth]{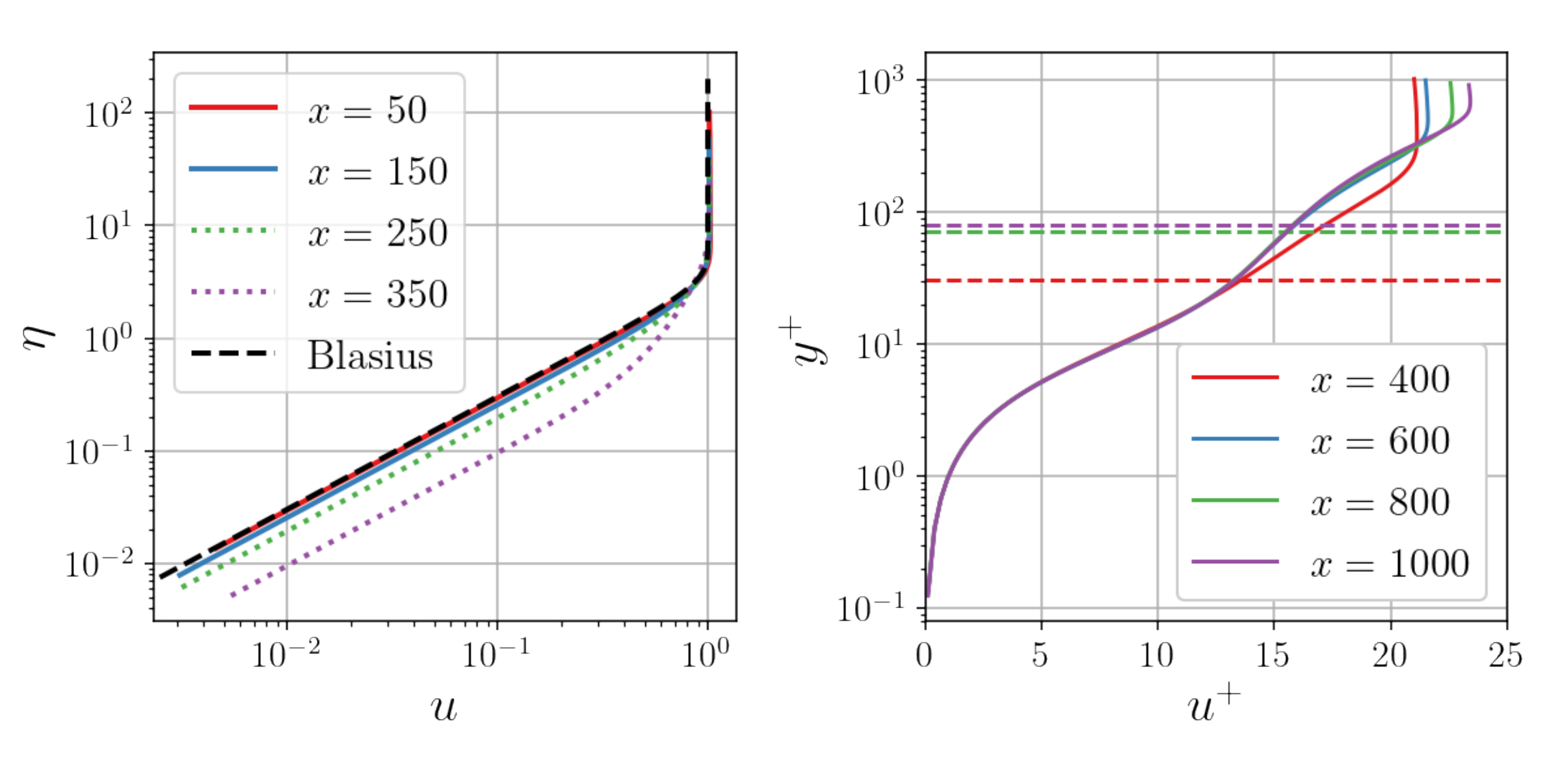}
	\end{overpic}
	\vspace{-.1in}
	\caption{
		The spatial dominant balance regimes are consistent with known self-similarity in the boundary layer.
		The laminar balance (Fig.~\ref{fig:boundary-layer}, purple and green) extends to approximately $x=200$, a region over which the mean profile approximately matches the Blasius solution (left).
		Similarly, the turbulent viscous sublayer (Fig.~\ref{fig:boundary-layer}, red) implies scaling with wall units (right).
		The mean profile with this scaling collapses until approximately the wall-normal extent of the identified viscous layer, as indicated by dashed lines.
	}
	\label{fig:self-similarity}
\end{figure}

Boundary layers are known to exhibit distinct self-similarity in both the laminar and turbulent regimes~\cite{Schlichting1955}.
In the laminar regime, Prandtl's boundary layer equations may be solved by introducing a similarity variable $\eta = y\sqrt{U_\infty/\nu x}$ and a dimensionless streamfunction $f(\eta)$ such that $u(x, y) = U_\infty f'(\eta)$.
With this ansatz the momentum equations reduce to the Blasius equation
\begin{equation}
2 f''' + f'' f = 0
\end{equation}
with boundary conditions $f(0) = f'(0) = 0$ and $f'(\infty) = 1$.
We expect that the mean field of the transitional boundary layer will be approximately given by the Blasius solution in the regions identified with negligible Reynolds stress (green and purple in Fig.~\ref{fig:boundary-layer}), which extend to approximately $x=200$.

The left panel of Fig.~\ref{fig:self-similarity} confirms this by comparing the numerically computed Blasius solution to the mean DNS profile.
The mean flow closely matches the Blasius profile for the region identified with a laminar dominant balance (solid lines), but significant discrepancies appear in the transitional region (dotted lines).

The scaling of the turbulent region is significantly more complicated, and aspects of it are still a topic of debate~\cite{Barrenblatt1997pnas,Zagarola1998jfm,Morrison2004jfm, Nickels2005prl,Marusic2010science, Smits2011arfm, Marusic2013jfm}.
In the inner layer, the balance between Reynolds stress and viscosity given by~\eqref{eq:viscous-sublayer} suggests the relevant length scale is determined by the friction velocity~\cite{Jimenez2013pof}.
In this case the appropriate scaling is in ``wall-units":
\begin{equation}
y^+ = \frac{y u_\tau}{\nu} \hspace{2cm} u^+ = \frac{u}{u_\tau}.
\end{equation}
The specific variation of $u^+$ with $y^+$ is usually throught to transition from a linear dependence in the very near-wall region to a the logarithmic ``law of the wall" before the scaling changes to that of the outer inertial sublayer.
Regardless of the specific functional form, the velocity profile should be self-similar in wall units throughout the viscous-dominated region (red in Fig.~\ref{fig:boundary-layer}).

The right panel of Fig.~\ref{fig:self-similarity} shows that the mean profile of the turbulent region does indeed collapse in wall units until approximately $y^+ \sim \mathcal{O}(10^2)$, as expected~\cite{Jimenez2013pof}.
The dashed lines indicate the wall-normal extent of the viscous layer; the self-similarity begins to noticeably deteriorate above this point for all streamwise locations.

The identified dominant balance regions are thus consistent with expected self-similarity in the boundary layer, in both laminar and turbulent regimes.
This behavior is not built into the algorithm in any way, but the result suggests that the method does indeed identify the correct physical balance in space.

\subsection{Optical pulse propagation}
\label{sec:results-optics}

\begin{figure}
	\centering
	\begin{overpic}[width=0.95\linewidth]{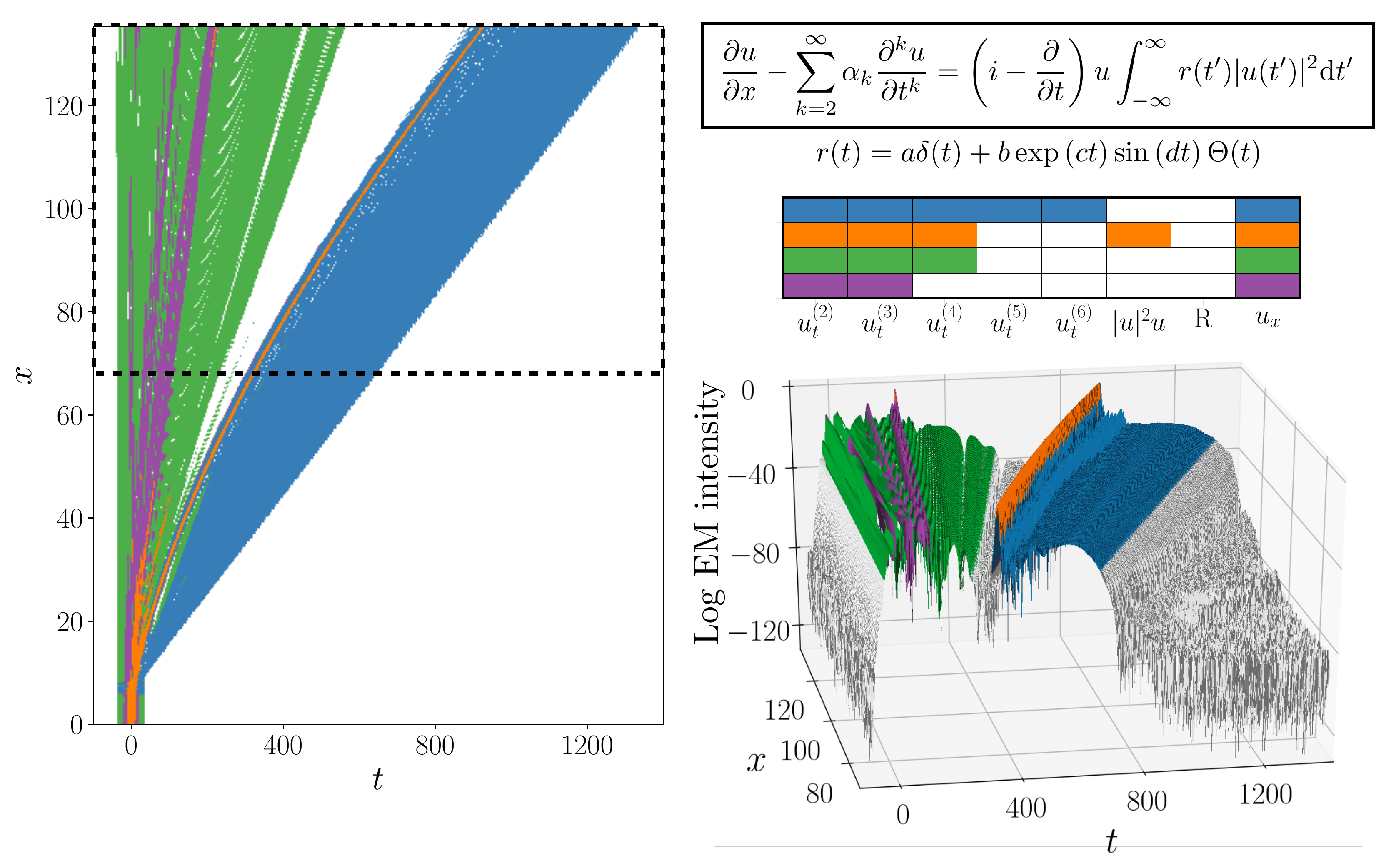}
	\end{overpic}
	\vspace{-.1in}
	\caption{  Identified balance models for the generalized nonlinear Schr\"{o}dinger equation. The governing equations are derived from Maxwell's equations in 1D with a nonlinear time-delayed polarization response.  Soliton propagation is understood to be maintained primarily by a balance between low-order dispersion and the cubic Kerr nonlinearity (delta-function component of the right-hand side integral) \cite{Mollenauer2006book}.  Although most of the field is identified with various linear dispersion relations, the strongest soliton is associated with cubic nonlinearity and dispersive terms through fourth order.  
	}
	\label{fig:optical-pulse}
\end{figure}

Another important example of dominant balance arises in nonlinear optics, where the interplay of an intensity dependent index of refraction with chromatic dispersion can generate localized optical solitons~\cite{Agrawal}.  The derivation of the governing evolution equations of the electric field envelope from Maxwell's equations shows that for ultra-short pulses of light (e.g. a few femtoseconds), the time response of the polarization field can yield~\cite{kutz2014solitons} a rich set of nonlinear dynamics.

Figure \ref{fig:optical-pulse} shows an example of a process known as supercontinuum generation, in which nonlinear processes act on a localized pulse of light to broaden the optical spectrum, stretching an initial 20-30~nanometer bandwidth to hundreds of nanometers.  This is typically accomplished in microstructured optical fibers~\cite{Dudley2010book}.
The governing equation in this case is derived from Maxwell's wave equation in one dimension through the rotating wave approximation and the slowly varying envelope approximation~\cite{kutz2014solitons}. 
The original PDE is linear and second order in a vacuum, but in order to handle complicated polarization responses in fibers the field is expanded about the frequency of the original pulse \cite{Blow1989, Mollenauer2006book}.
This ``center frequency" expansion leads to a Taylor series expansion of the linear polarization response, and the Raman convolution integral describing a time-delayed nonlinear response.

\paragraph{Governing equation}
The resulting PDE, known as a generalized nonlinear Schr\"{o}dinger equation (GNLSE) describes the evolution of the slowly varying complex envelope $u(x, t)$ of the pulse.  When nondimensionalized with soliton scalings \cite{Mollenauer2006book}, the envelope equation is
\begin{subequations}
	\begin{gather}
	\label{eq:gnlse}
	\pdv{u}{x} - \sum_{k=2}^{\infty} \alpha_k \pdv[k]{u}{t} = \left(i - \pdv{t}\right) u \int_{-\infty}^{\infty} r(t') | u(t') |^2 \dd t' \\
	r(t) = a \delta(t) + b \exp (ct) \sin (dt) \Theta(t).
	\end{gather}
\end{subequations}
The various constants ($\alpha_k$, $a$, $b$, $c$, $d$) describe the polarization response and are determined empirically.  

Although the spectral domain is often of practical interest for studies of supercontinuum generation, in the time domain the pulse exhibits soliton behavior, as shown in figure \ref{fig:optical-pulse}.
To leading order, the soliton propagation is typically understood to be maintained by a balance between the second order dispersion and the instantaneous part of the nonlinear response, or intensity-dependent index of refraction.
That is, evaluating the delta function component of the Raman kernel leads to the cubic Kerr nonlinearity.
If only this cubic nonlinearity and second order dispersion are retained, equation \eqref{eq:gnlse} is reduced to the usual nonlinear Schr\"{o}dinger equation (NLS):
\begin{equation}
\label{eq:nls}
i \pdv{u}{x} + \pdv[2]{u}{t} +  |u|^2 u =0.
\end{equation}

\paragraph{Identified dominant balance}
Figure \ref{fig:optical-pulse} shows the balance models obtained through the unsupervised balance identification procedure applied to regions of the field where the intensity is within 40 dB of the peak.
Most of the domain is associated with various linear dispersion relations, corresponding to different propagation speeds.
Only a narrow region containing the strongest soliton is identified with the instantaneous nonlinear response, suggesting that a linear description is sufficient for much of the domain.
The standard NLS equation is never identified, although the balance relation with cubic nonlinearity and fourth order dispersion (orange) is consistent with standard truncation of the linear response at third or fourth order \cite{Blow1989}.
Interestingly, the full Raman time-delay response is never selected as an important term, although this is understood to be a critical mechanism for the initial scattering.
Presumably the Gaussian mixture model approach is not sensitive enough to detect this, possibly due to the clearly invalid underlying assumption of normally distributed data.

To date, the ad hoc analysis of the various emergent structures have only qualitatively explained the origins of the observed phenomenon as the detailed numerical simulations do not disambiguate the contributions from the various terms of the high-fidelity model.  The dominant balance identification allows for a quantitative assessment of the emergent physics, even when solitonic structures are embedded in a sea of dispersive linear radiation.  Moreover, for the first time, the analysis suggests that the emergent solitons have a significant impact from 4th-order dispersion, as only recently discovered in pure-quartic soliton lasers~\cite{runge2020pure}.

\subsection{Geostrophic balance in the Gulf of Mexico}
\label{sec:results-geostrophy}

Geophysical fluid dynamics is a particularly complex field; a full description of ocean dynamics for instance requires not only the Navier-Stokes equations on a rotating Earth with complicated bathymetry, but must also account for the effects of varying salinity, temperature, and pressure via a nonlinear equation of state.
The ocean dynamics also couple to atmospheric and geological processes and solar forcing \cite{Gill1982book}.
Scaling analyses have been remarkably successful; despite the complexity of the dynamics, in many cases it can be argued that greatly simplified versions of the governing equations are sufficient to describe the dominant motions.

\paragraph{Governing equations}
Perhaps the most important model of this type is geostrophic balance.  To a first approximation, the surface currents can be modeled with the 2D incompressible Navier-Stokes equations on a rotating sphere:
\begin{subequations}
	\begin{align}
	\label{eq:rotating-nsU}
	& u_t + (\mathbf{u} \cdot \nabla) u + fv = -\frac{1}{\rho} p_x \\
	\label{eq:rotating-nsV}
	& v_t + (\mathbf{u} \cdot \nabla) v - fu = -\frac{1}{\rho} p_y,
	\end{align}
\end{subequations}
where $\rho$ is the density (in general a function of temperature, pressure, and salinity), and $x$ and $y$ are defined in the zonal and meridional directions, respectively.
The Coriolis parameter $f$ is given in terms of the Earth's angular velocity $\Omega$ and the latitude $ \phi $ by $ f= \Omega \sin \phi$.
Note that this equation already includes some approximations.  Compressibility, vertical motions, and both molecular and turbulent viscosities are all ignored in this model.
Nevertheless, these equations are a standard starting point for many analyses of large scale ocean dynamics.

\begin{figure}
	\centering
	\vspace{.2in}
	\begin{overpic}[width=\linewidth]{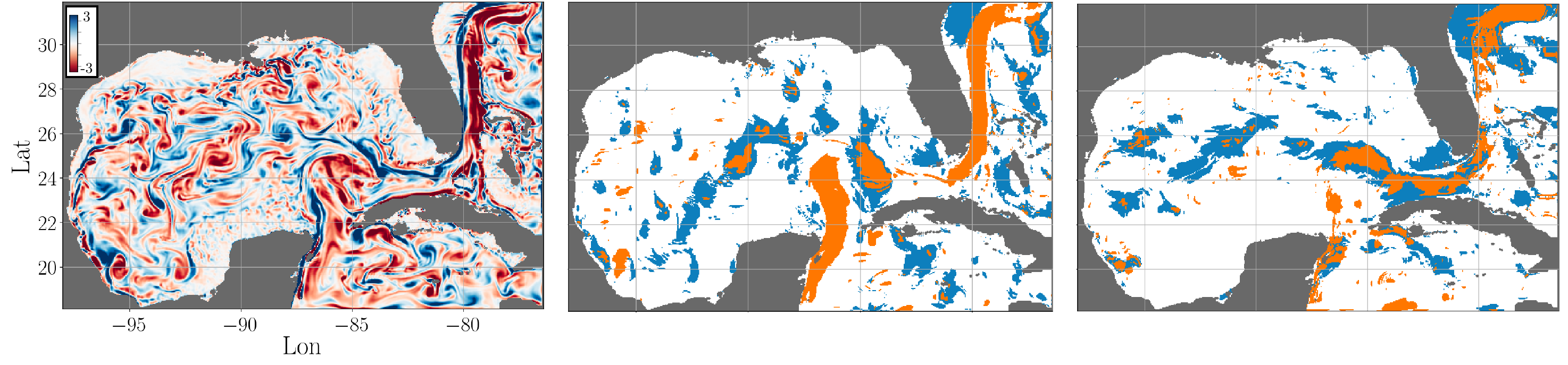}
		\footnotesize{
			\put(4, 24){\textbf{a}}
			\put(36, 24){\textbf{b}}
			\put(69, 24){\textbf{c}}
			\put(12, 24){Surface vorticity}
			\put(45, 24){Zonal balance}
			\put(76, 24){Meridional balance}
		}
	\end{overpic}
	\caption{
		Surface vorticity in the Gulf of Mexico along with identified balance models for zonal and meridional dynamics.
		Orange regions are identified with the geostrophic balance, while the red regions are time-varying in response to the pressure gradient and regions in light blue are associated with the linearized rotating Navier-Stokes equations.
	}
	\label{fig:geo-balance}
\end{figure}
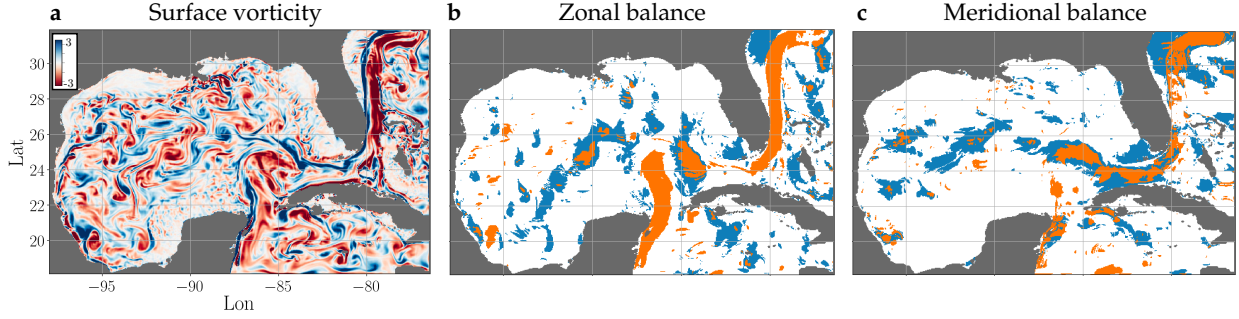

For flows with length scale $L$ and velocity scale $U$, the relative importance of the Coriolis terms compared to the inertial terms is given by the Rossby number, $ \mathrm{Ro} = U/ fL $.
In low Rossby number flows (relatively slow, large scale motions), the inertial terms become negligible and the dominant balance is between the Coriolis forces and pressure gradient forces:
\begin{subequations}
	\begin{align}
	& + fv = -\frac{1}{\rho} p_x \\
	& - fu = -\frac{1}{\rho} p_y.
	\end{align}
\end{subequations}
This balance is thought to describe most approximately steady large scale currents \cite{Gill1982book}.

\paragraph{Identified dominant balance}
We apply the unsupervised balance identification procedure to the high-resolution $1/25^\circ$ HYCOM reanalysis data for the Gulf of Mexico~\cite{hycom}.
In Fig.~\ref{fig:geo-balance} we explore the possibility of simultaneously identifying balance regimes in multidimensional systems.
Under the assumption that the same combinations of terms will appear in both zonal and meridional dynamics (but not necessarily in the same spatial regions), we compute the terms as usual and combine the equation spaces.
That is, the physically equivalent terms (pressure gradient, Coriolis forces, etc) are treated identically in the clustering process.
This simple approach would not generalize to multidimensional systems for which the constituent equations represent fundamentally different physics.
There is no such equivalence between the conservation equations for mass and energy, for example.

Figure~\ref{fig:geo-balance} shows the regions corresponding to balance models for this data.
The method identifies three regimes; geostrophic balance (orange), a balance between acceleration and the pressure gradient (red), and the linearized rotating Navier-Stokes equations (light blue).
The nonlinear advective term is not included in any of the models in this case, supporting the common use of linearized equations to study wavelike motions.
Geostrophic balance is primarily identified in regions corresponding to slow, large scale motions: the southern end of the Gulf Stream and the relatively stable current between Cuba and the Yucat\`{a}n Peninsula. 

Clearly the approximations in estimating gradients introduce significant error and variability into the balance identification procedure for this examples.
However, the identified models are consistent with the expected behavior according to classical arguments.
These results indicate some degree of robustness of the procedure and suggest that it may be applied to sufficiently clean experimental or data-assimilated observations.

\subsection{Generalized Hodgkin-Huxley model of an intrinsically bursting neuron}
\label{sec:results-neuron}

Networks of biological neurons in an animal's nervous systems communicate with each other through the propagation of electrical potentials.
These all-or-nothing events, known as \emph{action potentials} or  \emph{spikes}, are large deviations from the membrane electrical potential at rest, as measured between the inside and outside of a neuron.
Importantly, spikes can travel without significant degradation down the length of a neuron's long axon, which may be meters long. 

The celebrated Hodgkin-Huxley model for spiking neurons reproduces an action potential through a balance of currents from multiple ions, each of which moves through the cell's membrane across specialized channels and pores at different phases of a spike~\cite{hodgkin1952quantitative}. 
These non-linear partial differential equations were the first detailed biophysical model to quantitively describe the dynamic activity of neurons, and they underpin decades of ongoing attempts to understand more complex properties of neuronal electrical excitability~\cite{ermentrout2010mathematical}.

\paragraph{Governing equation}
The propagation of an action potential along an axon is well approximated by the cable equation of a cylinder of radius $a$,
\begin{align}
C_M \frac{\partial V}{\partial t} = \frac{a}{2 r_L} \frac{\partial^2 V}{\partial x^2} + \sum_{j} I_j,
\label{eq:hh_pde}
\end{align}
where $C_M$ is the membrane capacitance, $r_L$ is the resistivity inside the cell, and $I_j$ are each of the ionic currents in current per unit area due to the flow of ions into and out of the cell.

Hodgkin and Huxley originally modeled three (3) ionic currents: $I_{Na}$ sodium, $I_{K}$ potassium, and a leak $I_{L}$.
The dynamics of $V$ for a single action potential can then be expressed as a system of four (4) ordinary differential equations; the balance of currents in these equations reflect the biophysical mechanisms.

\begin{figure}
	\centering
	\begin{overpic}[width=0.75\linewidth]{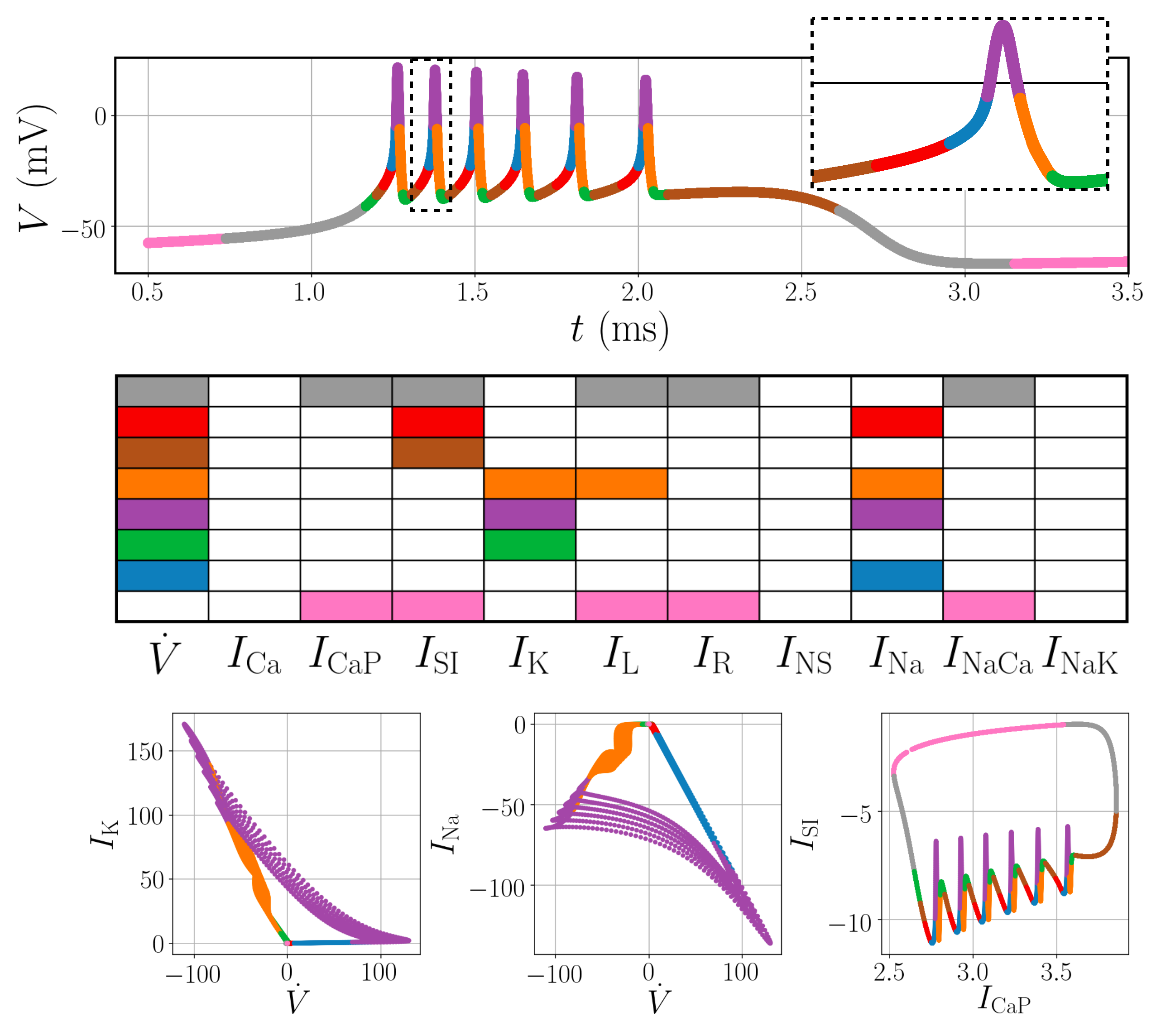}
	\end{overpic}
	\vspace{-.1in}
	\caption{
		Generalized Hodgkins-Huxley model for an intrinsically bursting neuron.
		Dynamics in quiescent periods are characterized by currents related to calcium concentration (pink and gray), while the spiking dynamics are dominated by the classic sodium-potassium cycle.
	}
	\label{fig:neuron}
\end{figure}

Adding more ionic currents and modeling the interactive balance of their dynamics produces more complex spiking behavior.
In particular, here we consider a generalized Hodgkin-Huxley model with ten (10) currents that simulates the intrinsically bursting pattern of spikes observed in the R15 neuron of the sea slug \emph{Aplysia}~\cite{canavier1991simulation}, as shown in Fig.~\ref{fig:neuron}.
The R15 neuron has been used to study the mechanisms underlying intrinsic bursting, where several action potentials are generated in rapid succession interspersed with relative quiet with constant inputs.
Under space-clamp conditions where an entire axon cable is considered to be spatially uniform, the equation describing the time-evolution of membrane voltage $V$ under applied external input $I_{\text{stim}}$ is
\begin{align}
C_M \dot{V} = - \sum_j I_j + I_{\text{stim}}.
\label{eq:hh}
\end{align}
Specifically, the ionic currents $I_j$ in our model are: $I_{Na}$ the fast sodium Na$^+$ current; $I_{Ca}$ the fast calcium Ca$^{2+}$ current; $I_{K}$ the delayed rectifier potassium current; $I_{SI}$ the slow inward calcium current; $I_{NS}$ the non-specific cation current; $I_{R}$ the anomalous rectifier current; $I_{L}$ the leakage rectifier current; $I_{NaCa}$ the sodium-calcium exchanger current; $I_{NaK}$ the sodium-potassium pump; $I_{CaP}$ the calcium pump.

\paragraph{Identified dominant balance}
Our dominant balance approach identifies several interpretable regimes of physics in the generalized Hodgkin-Huxley model that are largely consistent with known biophysics.
The addition of a set of calcium-dependent currents underly the slower oscillations between quiescence and excitable bursting, as evident in the slower limit cycle.
Notably, in these clusters, colored pink and gray in Fig.~\ref{fig:neuron}, the balance of ions is dominated by terms with strong calcium dependence ($I_{CaP}$, $I_{SI}$, and $I_{NaCa}$).
In contrast, the time-course of $V$ at each fast spike is dominated by voltage-gated ionic currents.
In Fig.~\ref{fig:neuron}, the rising part of each spike is mediated by activation of sodium channels, and the inward $I_{SI}$ and $I_{Na}$ increase $V$ (red and blue).
$V$ reaches peak voltage as the sodium channels inactivate and delayed rectifier potassium channels $I_{K}$ activate (purple). The exit of potassium from the cell decreases $V$ back towards the resting potential.

There are three currents that have not been identified to belong to any cluster: the fast calcium current, sodium-potassium pump, and the non-specific cation current.
Although these are dynamically important for the model, they are relatively small compared to the other terms ($\mathcal{O}(0.1-1)$ compared to $\mathcal{O}(100)$ for the spiking dynamics) and so they don't appear to participate in any of the local dominant balance relationships identified by this method.
This is a similar situation to the Raman time-delay nonlinearity in the optical pulse propagation example (Sec. \ref{sec:results-optics}) and the nonlinear advection in the Gulf of Mexico (Sec. \ref{sec:results-geostrophy}).
In all of these cases, the influence of the neglected terms appears to be of a more subtle nature than the dominant balance physics we explore in this work.

\subsection{Rotating Detonation Analog System}

\begin{figure}
	\begin{center}
		\begin{overpic}[width=0.9\linewidth]{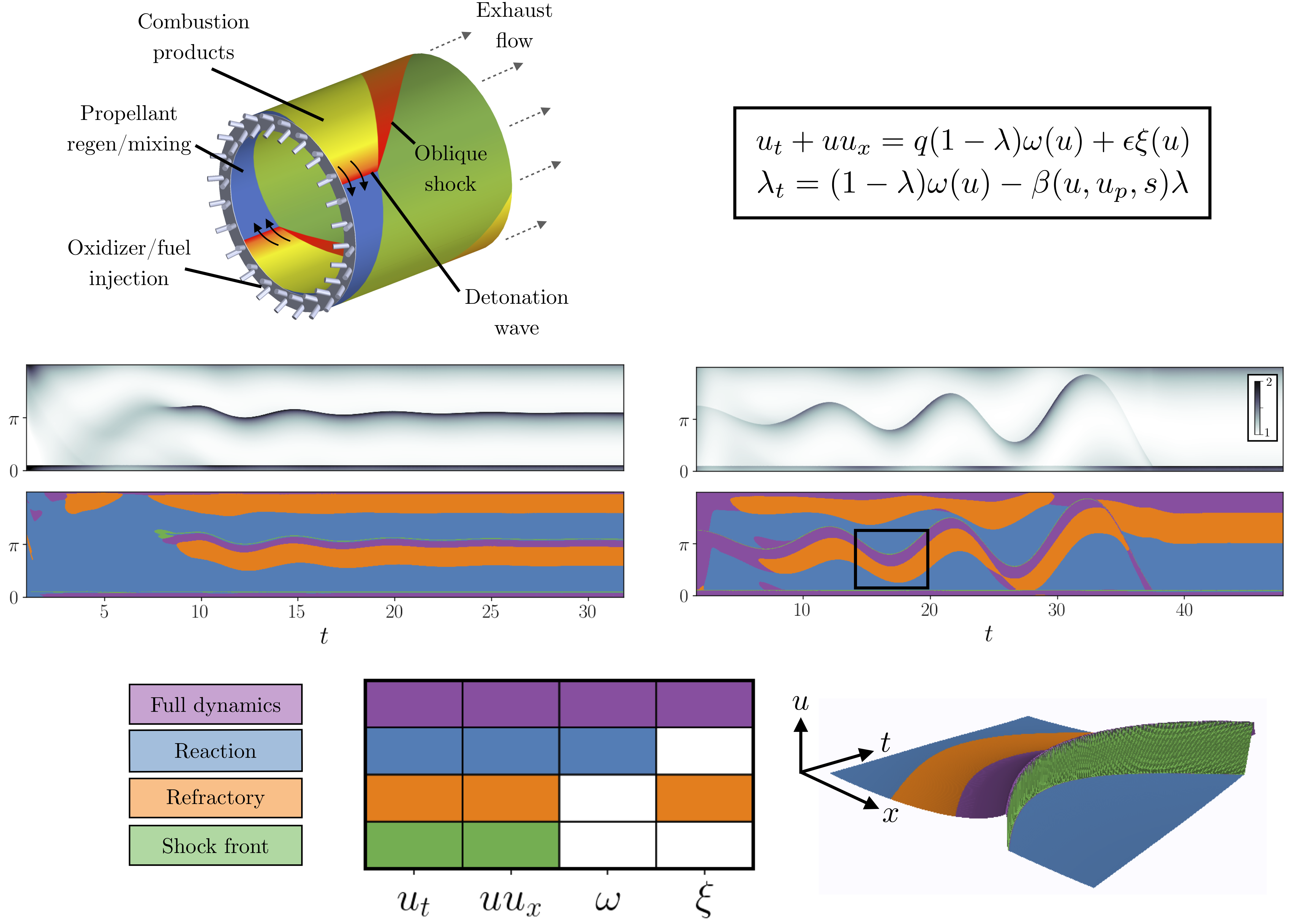}
		\end{overpic}
		\vspace{-.2in}
	\end{center}
	\caption{ Model of combustion dynamics in a rotating detonation engine.
		The dynamics on the thin shock front are determined by the canonical Burgers balance (green), followed by activation of the gain and loss terms (purple).
		Following the combustion front, the balance transitions to the refractory exhaust-dominated period (orange). 
		The rest of the domain is characterized by a combination of the Burgers dynamics with background energy input.
	}
	\vspace{-.2in}
	\label{fig:rde}
\end{figure}

The \textit{Rotating Detonation Engine} (RDE) is a novel rocket engine combustor configuration that exploits the self-steepening properties of reactive compressible flows in confined, periodic geometries (such as an annular chamber, as depicted in Fig. \ref{fig:rde}) to form traveling detonation waves that persist in time~\cite{Lu2014}. For an annular RDE, fuel and oxidizer is injected at the head-end of the device, where the propellant streams rapidly mix to form a detonable mixture. Once ignited (with a spark plug, for example), the rapid exothermal chemical reaction induce gradients in temperature and density within the flowfield. Because chemical kinetics are accelerated with increases in temperature, the reaction front can self steepen, eventually forming a supersonically traveling shock wave coupled to a region of rapid chemical heat release. This shock-reaction structure - the detonation wave - can travel about the annular combustion chamber so long as the ingested propellant mixture (i) contains enough chemical potential to offset dissipative effects (such as rapid expansion of the flow downstream and heat transfer to the engine walls) and (ii) is well mixed~\cite{Bykovskii2006,Zhdan2007,Yamada2010,Nordeen2015}. Thus, this combustor configuration features prominently detonative heat release - a departure from conventional constant-pressure deflagration-based engines used in aerospace applications. The potential advantages of the RDE over its deflagration-based counterparts include a potentially greater thermodynamic cycle efficiency, greater power density, and significant mechanical simplification. 

The behavior of the RDE, including the development of the flowfield and the behavior of the detonation waves, is a consequence of the intricate coupling of the multi-scale physical processes present in the system~\cite{Koch2020pre}. The fundamental processes - injection, mixing, combustion, and exhaustion - all posses unique time and spatial scales. These range from the thickness and speed of the detonation shock front (the shortest and fastest present in the RDE system, respectively) to the length of the combustor and associated residence time of a fluid particle (the longest scales present). However, despite these scales varying by several orders of magnitude, they are intimately coupled: behind the detonation wave, the combustion products must be ejected within the period of the wave. Similarly, new propellant must be injected and sufficiently mixed prior to wave arrival. By varying the associated scales of these physics, a variety of operating modes can be obtained, including various wave counts, directions, speeds, and pattern formation with counter-propagating waves.

\paragraph{Phenomenological Model}
A phenomenological model was recently proposed \cite{Koch2020pre} to relate the associated scales of the fundamental processes of the RDE. The model adapts the inviscid Burgers' equation to a periodic domain with imposed gain (energy input from combustion) and dissipation (exhaust processes). The evolution of a representative quantity $u(x,t)$ is supplemented with a evolution equation for a combustion progress variable, $\lambda(x,t)$, which describes the balance of gain depletion and gain recovery. The model is given as:

$$
u_t + u u_x = q (1 - \lambda) \omega(u) + \epsilon \xi(u)
$$
\begin{equation} \label{eq:rde}
\lambda_t = (1-\lambda)\omega(u) - \beta(u,u_p,s)\lambda .
\end{equation}
where $u(x,t)$ is analogous to an intensive property of the fluid, such as internal specific energy, $q$ is the energy release associated with the reactive mixture, $\omega(u)$ is the submodel for kinetics, $\xi$ is the submodel for exhaust (with a loss coefficient $\epsilon$), and $\beta(u,u_p,s)$ is the injection and mixing submodel with parameters for an injection sensitivity cuttoff $u_p$ and overall timescale $s$. For the presented cases, the submodels for kinetics, dissipation, and injection are unchanged from the original presentation of the model in \cite{Koch2020pre}. The model has been shows to qualitatively reproduce the nonlinear dynamics of the collection of detonation waves present in an RDE, including wave nucleation, destruction, modulation, and mode-locking.

\paragraph{Identified dominant balance}
Two simulations of Eq. \ref{eq:rde} are shown in Fig. \ref{fig:rde} in the wave-attached reference frame, each showing a canonical bifurcation of number of detonation waves present in the system. Application of our dominant balance method identifies four interpretable and distinct regions of physics, as shown in Fig. \ref{fig:rde}. We first examine these balances within the context of a steadily propagating wave. At the front of the wave is a thin region shaded in green. This region corresponds to the shock physics of the classic Burgers' equation. For this region specifically, $\omega(u)$ is approximately negligible, as the kinetics - an exponential function of $u$ for this case - are slow until $u$ can activate $\omega(u)$. Indeed, shortly behind the shock front appears a region shaded in blue that indicates appreciable energy release into the system. This region is also relatively thin: an accumulation of $u$ inside the domain is required before the nonlinear dissipation submodel - a quadratic function of $u$ - becomes significant. This occurs in the purple shaded region, where the rate of energy input to the system (which is now slowed because of the $(1-\lambda)$ multiplier with $1>\lambda>>0$) is of the same order as the dissipation term. Once $\lambda \approx 1$, energy input becomes negligible, though dissipation is still significant; this region is shaded in orange. This region constitutes the \textit{refractory} period behind the detonation wave where $u$ and $\lambda$ approach rest values. For the presented simulations, the remainder of the domain is characterized by the balance of the nonlinearity of the medium (Burgers' flux) and background energy input. 

In Fig. \ref{fig:rde}a, after an initial start-up transient for a single wave, a second wave nucleates and \textit{mode locks} with the other established wave. This case highlights the result in a shift in balance physics. Prior to time $t = 7.5$, the background energy input occurring in the large blue region allowed for the accumulation of $u$ inside the domain, thereby accelerating the rate of energy input. The positive feedback loop of energy accumulation and rate of energy input is unchecked until the dissipation term offsets the energy input term. Thus, at time $t=7.5$, seen is the formation of a new shock front behind which is a coupled reaction and refractory zone. Immediately after nucleation, the asymmetry of wave positions causes an imbalance of available $\lambda$ to each of the waves. For this case, after oscillations in phase difference between the waves, they mode-lock to the same velocity and separation distance.

A similar wave asymmetry is present in the simulation of Fig. \ref{fig:rde}b. Here, two waves briefly co-exist in a domain where there exists only enough energy flux to support a single wave. The initial perturbation in wave separation grows exponentially, until one wave overruns the other in the final high-amplitude oscillation of phase difference. During this period of instability, observed are similar oscillations in the thicknesses of regions of dominant balances. Perhaps more noteworthy, however, are (i) the apparent phase shift of these oscillations relative to the wave positions and (ii) the role of the refractory period in wave destruction. Each of the identified dominant balance regions similarly possesses an oscillatory thickness. Each layer has a unique phase shift behind the detonation wave to which the layer is attached. In the final oscillation of the two waves, one wave enters the refractory region of the other. The effect upon the wave is dramatic: it immediately weakens (lower amplitude and slower speed) and sheds its own refractory region. Therefore, once the opposing wave encounters this weak wave (with no significant refractory region), the weak wave is easily overtaken and destroyed. The remaining wave propagates stably after this bifurcation.

\section{Discussion}
\label{sec:discussion}

In one guise or another, dominant balance analysis has played a major role in the development of our understanding of many complex systems.
In this paper, we have proposed a method of identifying dominant balance regimes in an unsupervised manner directly from data.
This approach leverages our understanding of the full physical complexity in the form of governing equations, but by using simple clustering and sparse approximation methods, we avoid any a priori assumptions about balance relations.
Nevertheless, the method identifies dominant balance relationships that either recover classical scaling analysis (in the case of the boundary layer and Gulf of Mexico) or confirm arguments based on physical intuition (in the case of nonlinear optics, the Hodgkin-Huxley model, and the combustion analog).

The critical step in this process is the equation space perspective.
By considering each term in the governing equation to describe a direction in this space, the dominant balance relations naturally manifest via restriction to subspaces, dramatically reducing variance in directions corresponding to negligible terms.
This observation enables the Gaussian mixture model to identify clusters with variance in different directions, and the sparse principal components analysis to extract sparse subspaces by finding directions with significantly nonzero variance.
These machine learning tools are applied in a targeted and clearly motivated context, and the equation space perspective necessarily ties the output to underlying physics.

This data-driven approach has the same goal as traditional methods such as scaling analysis, but introduces several new features.
It is a principled, objective approach that does not require the assumption of asymptotic parameter regimes, while providing an estimate of the locally active physical processes throughout domains with arbitrarily complex geometries.
The proposed method retains the advantages of the classic approach, but generalizes to a range of disciplines to which traditional analysis cannot readily be applied.

Dominant balance analysis has historically been a critical tool for understanding local physical behavior in complex systems.
Non-asymptotic data-driven methods could be used to better understand the behavior of more exotic dynamics such as non-Newtonian turbulence~\cite{Samanta2013pnas}
or to study important transitional behavior in cases where the asymptotics are already well known~\cite{Hof2004science, Eckhardt2007arfm,Avila2011science}.
In the latter case, a clear understanding of the active mechanisms has proven crucial to successful control strategies~\cite{Du2000science,Hof2010science}.

The existence of dominant balance limiting regimes even in complex nonlinear spatiotemporal systems is consistent with the observation that these systems can often be described with sparse representations in function space~\cite{Brunton2016pnas, Rudy2017sa}.
Building on this insight, we may even be able to identify local dominant balance behavior in spatiotemporal systems without clear governing equations, such as neuroscience, epidemiology, ecology, active fluids, and schooling.
For example, the inclusion of spurious terms in the governing equation can be readily detected in the equation space representation (see Supplementary Information); in future work this feature might be leveraged to identify local balance relations in the absence of global conservation equations.
This approach thus stands to shed light on more exotic physical processes that have remained elusive to traditional analysis. 

However, as with all applications of machine learning and data science methods to physical systems, a critical step in application to any system will be careful validation that the balance identification procedure reproduces the expected results.
The dominant balance modeling approach described here is designed to build on, rather than circumvent, physical expertise.
The study of dominant balance regimes has been foundational to our understanding of many complex systems; we hope that data-driven methods can integrate with this legacy to enable even wider applicability.

\vspace{-0.2cm}
\section*{Acknowledgements}
\vspace{-0.2cm}
JLC acknowledges support from the NDSEG fellowship. 
JNK acknowledges support from the Air Force Office of Scientific Research (AFOSR FA9550-17-1-0329). 
BWB acknowledges support by the Washington Research Foundation. 
SLB acknowledges funding support from the Air Force Office of Scientific Research (AFOSR FA9550-18-1-0200) and the Army Research Office (ARO W911NF-19-1-0045).  
The authors also acknowledge support from the Defense Advanced Research Projects Agency (DARPA PA-18-01-FP-125). 

\vspace{-0.2cm}
\section*{Appendix A: Data provenance}
\vspace{-0.2cm}

\paragraph*{Direct numerical simulation of flow past a circular cylinder.}
We simulate this configuration at $\Re=100$ with unsteady incompressible DNS using the open source spectral element solver Nek5000 \cite{Nek5000}.  
The domain consisted of 17,432 seventh order spectral elements 
on $ x, y \in (-20, 50) \times (-20, 20) $, refined close to a cylinder of unit diameter centered at the origin.  Diffusive terms are integrated with third order backwards differentiation, while convective terms are advanced with a third order extrapolation.
The results of this simulation have been validated against those of the immersed boundary projection method \cite{Taira2007} by comparing aerodynamic coefficients and vortex shedding frequency.
We extract the vorticity field and spatial terms in equation \eqref{eq:2d-ns} directly from the solver for further analysis.  Time derivatives for dominant balance identification were estimated with a second order central difference.

\paragraph*{Direct numerical simulation of a transitional boundary layer.}
To study dominant balance physics in the turbulent boundary layer, we use the transitional DNS by Lee and Zaki \cite{Zaki2013, Lee2018, Wu2019prf}, openly available from the Johns Hopkins Turbulence Database \cite{Perlman2007jhtdb, Li2008jhtdb}\footnote{https://doi.org/10.7281/T17S7KX8}.
The full computational domain consists of a long flat plate with an elliptical leading edge.
The extent of the domain (in units defined by the plate half-thickness) is $(x, y, z) \in (1040, 40, 240)$ with periodic boundary conditions in the spanwise ($z$) direction, discretized to $(Nx, Ny, Nz) = (4097, 257, 2049)$.
Since the configuration of interest is a zero pressure gradient flat plate boundary layer, the DNS results are only saved once the flow passes the elliptical leading edge ($x > 30.2185$).
The inflow consists of small amplitude free-stream turbulence superimposed on a uniform streamwise velocity $U_\infty$ incident on the plate.
The interactions of these perturbations with the laminar boundary layer cause a downstream transition to turbulence \cite{Zaki2013}.

Since we are interested here in the mean momentum balance, we only use the 2D mean field (also available from JHTDB), which was computed from 4701 data snapshots once the flow reached a statistically stationary state.
Without direct access to the gradients, we compute the constituent terms of the RANS equations with second-order accurate finite differences, as shown in Fig. \ref{fig:boundary-layer}b.
Although some of these fields show small fluctuations, the overall smoothness suggests the statistics are approximately converged.

\paragraph*{Supercontinuum generation in photonic crystal fiber.}
The generalized nonlinear Schr\"{o}dinger equation (GNLSE), nondimensionalized with soliton scaling \cite{Mollenauer2006book}, is given by Eq. \eqref{eq:gnlse}.
The various constants describe the polarization response and are determined empirically.  In this case we use the values described by Dudley \textit{et al} for photonic crystal fiber \cite{Dudley2010book}.  We also use the split-step spectral method and initial conditions described in these works to simulate the pulse propagation\footnote{MATLAB code freely available at http://www.scgbook.info/}.

\paragraph*{Surface currents in the Gulf of Mexico.}
We study the high-resolution $1/25^\circ$ HYCOM reanalysis data for the Gulf of Mexico~\cite{hycom}.
We use data from only the first field in the data set, corresponding to January 1993.  
Data-assimilated fields are available for the 2D velocity components, sea surface temperature, salinity, and sea surface height; vorticity is shown in Fig.~\ref{fig:geo-balance}.

We must therefore estimate time derivatives and both velocity and pressure gradients to compute the terms in Eqns.~\eqref{eq:rotating-nsU} and~\eqref{eq:rotating-nsV}.
Since this information is not directly accessible from the model (as for the numerical examples), we use finite differences to estimate the velocity derivatives.
The pressure field itself is also not available; as a rough estimate we use the residuals of the left-hand side of Eqns.~\eqref{eq:rotating-nsU} and~\eqref{eq:rotating-nsV} in place of pressure gradients.
We also assume constant density throughout the field.
Finally, since this field is two-dimensional but the terms in each evolution equation represent the same physics, we simply stack the features for each velocity component into a single $ (2N \times 4 ) $ matrix with columns corresponding to acceleration, convection, Coriolis forces, and the pressure gradient.
Although these are strong assumptions and approximations, we would expect them to only make the dominant balance identification problem more difficult, since they represent attempts to deal with limited information about the system.

\paragraph*{Generalized Hodgkins-Huxley model of a bursting neuron.}

A full set of model equations, including biophysical parameters, follow \cite{canavier1991simulation} and are given in the simulation code.
Briefly, gating variables following Hodgkin-Huxley form are described by solutions to differential equations of the general form
$\dot{z} = (z_{\infty} - z) / \tau_z$,
where $z_{\infty}$ are the steady-state values and $\tau_z$ are the time constants associated with the gating variable $z$.
To produce the data used in our analysis, this system of ordinary differential equations was integrated numerically in MATLAB using ode15.

\subsection*{Rotating Detonation Analog}
The system was simulated using the open-source PyClaw package for hyperbolic eqations with the same parameters as~\cite{Koch2020pre}.
The governing equations are
\begin{subequations}
	\begin{equation}
	u_t + u u_x = q (1 - \lambda) \omega(u) + \epsilon \xi(u)
	\end{equation}
	\begin{equation} 
	\lambda_t = (1-\lambda)\omega(u) - \beta(u)\lambda .
	\end{equation}
\end{subequations}
Following the original paper, we use
\begin{subequations}
	\begin{equation}
	\xi(u) = u^2
	\end{equation}
	\begin{equation}
	\omega(u) = \exp \left[ \frac{u - u_c} {\alpha } \right]
	\end{equation}
	\begin{equation}
	\beta(u) = \frac{s}{ ( 1 + e^{k(u - u_p)} }
	\end{equation}
\end{subequations}
with parameters as follows:
\begin{center}
	\begin{tabular}{||c c c c c c||} 
		\hline
		$q$ & $\epsilon$ & $u_c$ & $\alpha$ & $k$ & $u_p$ \\ [0.5ex] 
		\hline\hline
		1.0 & 0.11 & 1.1 & 0.3 & 5 & 0.5 \\ [1ex] 
		\hline
	\end{tabular}
\end{center}
Nondimensional time $\tau$ is given in terms of the Chapman-Jouguet wave speed and domain size, in this case $\tau = t/\pi$.
The nucleating wave is generated with an initial condition of $u(x, 0) = 1.5 \sech^10 (x)$ and $s=3.5$, while the annihilating waves are generated from an initial condition of two sech pulses and $s=2$.

\vspace{-0.2cm}
\section*{Appendix B: Parameter tuning}
\vspace{-0.2cm}

The proposed method was designed to minimize the number of hyper-parameters that need to be tuned.
However, there are two important parameters that must be selected: the number of clusters for the Gaussian mixture model (GMM), and the $\ell_1$ regularization for sparse principal components analysis (SPCA).

Since the data is not actually drawn from a mixture of Gaussian distributions it can be difficult to make a principled choice for the number of GMM clusters.
Intuitively, if there are too few clusters the GMM procedure cannot be expected to capture all of the distinct directions of variance in the equation space.
The secondary SPCA reduction makes the method somewhat robust to this parameter; the final balance models tend to be similar provided that there are enough clusters.
However, if there are too many clusters, the constituent distributions of the mixture model may not contain enough points to be dominated by a single principal component.

The $\ell_1$-regularization for SPCA is somewhat easier to choose with a simple model selection procedure.
A larger regularization value tends to yield more sparsity in the leading principal component, corresponding to neglected terms in the cluster.
We define the residual for a given regularization value as the $\ell_2$-norm of the neglected terms across all clusters.
For example, if SPCA with a regularization of 0.1 yields a principal component with a zero in the direction corresponding to viscosity for one of the clusters, the SPCA residual for 0.1 in that cluster would be the magnitude of the viscous terms in that cluster.
Sweeping a range of regularization values yields a Pareto-type curve showing the tradeoff of sparsity against descriptiveness.

\begin{figure}
	\begin{center}
		\begin{overpic}[width=0.95\linewidth]{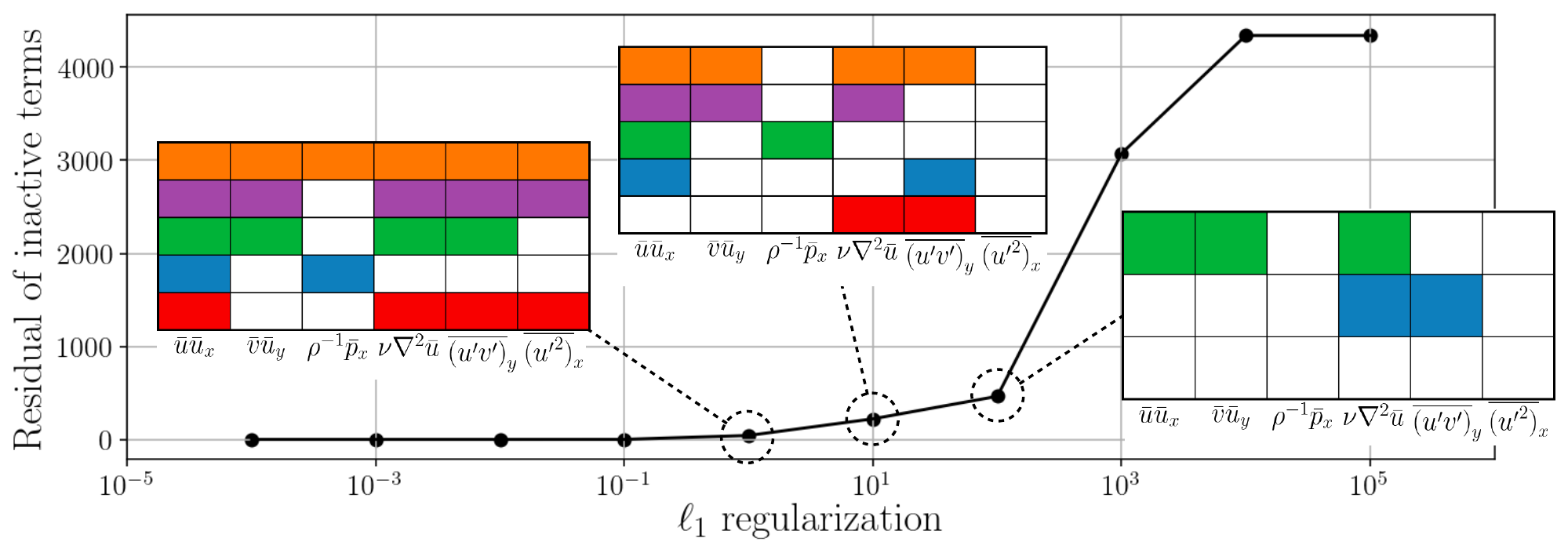}
		\end{overpic}
		\vspace{-.2in}
	\end{center}
	\caption{
		Model selection procedure used to choose a sparse regularization value for the principal components analysis, demonstrated on the turbulent boundary layer example.
		Although there is some flexibility depending on the desired accuracy and simplicity in the specific application, the residual of neglected terms suggests a range of appropriate values.
		In this work we chose regularizations that were as sparse as possible but spanned most of the original terms in the equation and had relatively small residuals (middle panel).
		Often this led to a set of balance relations, each with 2-3 terms, which collectively captured much of the richness of the full system.
	}
	\label{fig:regularization}
\end{figure}

This metric offers a guideline for choosing an appropriate regularization, although there is still some flexibility in the specific value.  As Fig. \ref{fig:regularization} shows, tuning the regularization differently yields a different set of balance models.
As with many model selection procedures, a different value may be selected depending on the desired level of descriptiveness and parsimony.
Based on physical considerations, in this work we looked for regularizations that resulted in a diversity of balance relations with 2-3 active terms each (middle panel of Fig. \ref{fig:regularization}).

\vspace{-0.2cm}
\section*{Appendix C: Model uncertainty}
\vspace{-0.2cm}

The idea of dominant balance is not necessarily clearly defined outside of asymptotic regimes; strictly speaking, all terms in a model are likely to have some nonzero contribution throughout the domain of interest.
Considering for example the cylinder wake, clearly the boundary layer is not steady, nor is the far-field region actually inviscid.

\begin{figure}
	\begin{center}
		\begin{overpic}[width=0.85\linewidth]{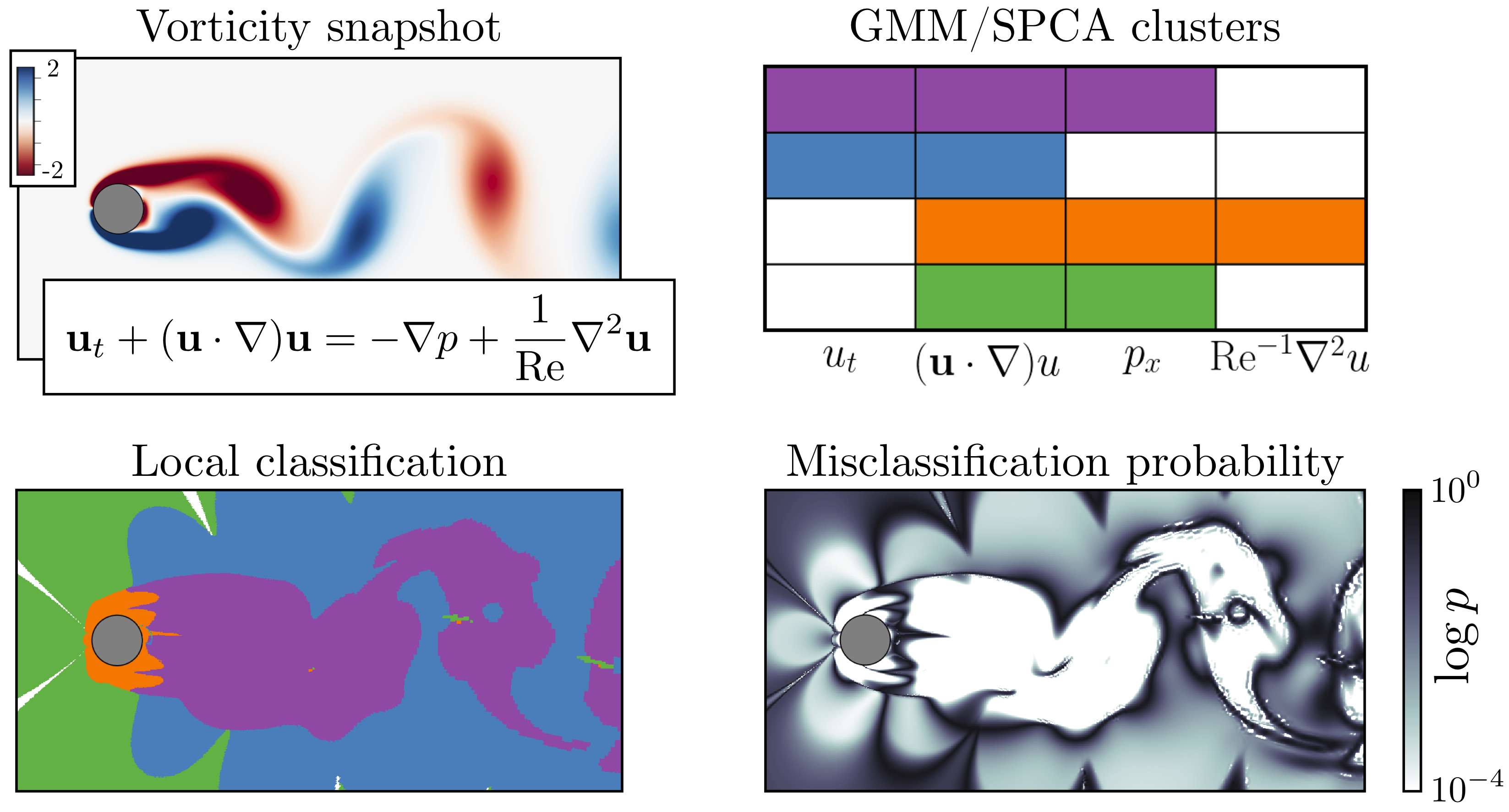}
		\end{overpic}
		\vspace{-.2in}
	\end{center}
	\caption{
		Uncertainty estimation for the dominant balance identification procedure.
		The Gaussian mixture model clusters points in the domain by assigning a probability of belonging to each Gaussian distribution.
		Summing the probabilities that each point belongs to a GMM cluster which SPCA reduces to the same balance model gives an overall estimate of the uncertainty associated with the identified dominant balance.
	}
	\vspace{-.2in}
	\label{fig:uncertainty}
\end{figure}

Fortunately, since GMM is a probabilistic clustering method it comes with a natural notion of uncertainty.
The clustering procedure assigns to each point a probability of belonging to each cluster.
We can propagate this through the SPCA reduction by summing the probabilities that each point in the field belongs to one of the clusters that reduces to the same balance model.
This results in an estimate of the probability of misclassification of each point, as shown in Fig. \ref{fig:uncertainty}.
As expected, this measure generally becomes large in transitional regions.
However, keeping their approximate nature in mind, the balance models offer a principled and intuitive segmentation of the domain according to the dominant physics.

\newpage
\setlength{\bibsep}{3pt plus .5ex}
\begin{spacing}{.01}
	\small
	\bibliographystyle{unsrt}
	\bibliography{refs}
\end{spacing}

\end{document}